\documentclass[%
 reprint,
 amsmath,amssymb,
 aps,
prx,
longbibliography]{revtex4-1}

\usepackage{graphicx}
\usepackage{dcolumn}
\usepackage{bm}

\usepackage{color} 
\usepackage{tabularx}

\begin{document}

\preprint{APS/123-QED}

\title{Finite Dissipation in Anisotropic Magnetohydrodynamic Turbulence}

\author{Riddhi Bandyopadhyay$^1$}

\author{S. Oughton$^2$}

\author{M. Wan$^3$}

\author{W. H. Matthaeus$^{1,4}$}
\email{whm@udel.edu}

\author{R. Chhiber$^1$}

\author{T. N. Parashar$^{1,4}$}

\affiliation{$^1$Department of Physics and Astronomy, University of Delaware, Newark, DE 19716, USA}
\affiliation{$^2$Department of Mathematics and Statistics, University of Waikato, Hamilton 3240, NZ}
\affiliation{$^3$Department of Mechanics and Aerospace Engineering, Southern University of Science and Technology, Shenzhen, Guangdong 518055, People's Republic of China}
\affiliation{$^4$Bartol Research Institute, University of Delaware, Newark, DE 19716, USA}
\date{\today}

\begin{abstract}
	
In presence of an externally supported, mean magnetic field a turbulent, conducting medium, such as plasma, becomes anisotropic. This mean magnetic field, which is separate from the fluctuating, turbulent part of the magnetic field, has considerable effects on the dynamics of the system. In this paper, we examine 
the dissipation rates for decaying incompressible magnetohydrodynamic (MHD) turbulence
with increasing Reynolds number, and 
in the presence of a mean magnetic field of varying strength. 
Proceeding numerically, 
we find that 
as the Reynolds number increases, the dissipation rate 
asymptotes to a finite value for each magnetic field strength, confirming the 
K\'arm\'an-Howarth hypothesis as applied to MHD. 
The asymptotic value of the dimensionless dissipation rate is initially 
suppressed from the zero-mean-field value by the mean magnetic field but then approaches a constant value for higher 
values of the mean field strength. Additionally, for comparison, we perform a set of two-dimensional (2DMHD) and a set of reduced MHD (RMHD) simulations. We find that the RMHD results lie very close to the values corresponding to the high mean-field limit of the three-dimensional runs while the 2DMHD results admit distinct values far from both the zero mean field cases and the high mean field limit of the three-dimensional cases. 
These findings provide firm underpinnings for 
numerous applications in space and astrophysics wherein von K\'arm\'an decay
of turbulence is assumed. 
\end{abstract}

\maketitle
\section{Introduction and Background}
Turbulence is a ubiquitous, although incompletely 
understood phenomena. 
In turbulent astrophysical plasmas, velocity 
and magnetic fields are often equally important, and magnetohydrodynamic (MHD)
turbulence, the case considered here, becomes an appropriate description. 
Energy supply for turbulence 
usually originates at large scales due to 
some type of stirring or driving mechanism,
after which 
the energy cascades to smaller scales, 
finally reaching 
the dissipation scale where the energy is dissipated into heat by viscosity and resistivity. 
For incompressible hydrodynamics, the single scalar viscosity 
$\nu$ parameterizes microscopic nonideal effects, such that, for 
laminar flows the energy dissipation rate vanishes when 
$\nu \to 0$. However, for turbulent flows, following the hypothesis
stated by Taylor~\cite{Taylor1935} and von K\'arm\'an \& Howarth
~\cite{Karman1938PRSL} 
and later employed by Kolmogorov~\cite{Kolmogorov1941c},
in the zero viscosity, infinite Reynolds number $Re \sim 1/\nu \to \infty$
limit, the energy dissipation rate approaches a 
constant, nonzero limit\cite{Donzis2005JFM}. 
(The renowned Kolmogorov theory of universal scaling in the 
inertial range (K41)~\cite{Kolmogorov1941a}, in essence assumes this 
limiting behavior.)
This so-called
``dissipative anomaly'' is well supported in experiments and computations \cite{Ravelet2008JFM,SaintMichel2014PoF}. 

For the case
of incompressible MHD, there are two relevant dissipation coefficients, viscosity $(\nu)$ and resistivity $(\mu)$, associated with velocity and magnetic field, respectively. The limit of zero viscosity and resistivity $(\nu, \mu \rightarrow 0)$ is again of interest with regard to fully developed turbulent systems. 
 In practice one finds that, for essentially all astrophysical as well as terrestrial turbulent systems, the mechanical and magnetic Reynolds numbers (inverse of viscosity and resistivity in non-dimensional units) are of large, and sometimes
colossal magnitude, justifying the success of turbulence phenomenologies like K41. 
For example, the effective 
mechanical and magnetic Reynolds number, 
in the solar wind, are of the order of $10^{5}$ or more \cite{Matthaeus2005PRL} (also see \cite{Verma1996JGR}), 
and much larger in the interstellar medium.
Therefore, one might be tempted to assume $\nu, \mu = 0$ as an approximation. However,  in analogy to the hydrodynamics case, 
the limit $\nu, \mu \rightarrow 0$ is very different from the $\nu, \mu = 0$ case. This is manifested in the counterintuitive phenomena that as the limit $\nu, \mu \rightarrow 0$ is approached, the total turbulent energy dissipation rate does not vanish, but remains finite. MHD energy dissipation is exactly zero for 
the ideal case $(\nu, \mu = 0)$. Onsager \cite{Onsagar1949d} conjectured that this problem of energy-dissipation anomaly arises due to lack of smoothness in the velocity field increments in the context of three-dimensional hydrodynamic turbulence. Based on this idea of lack of smoothness in the velocity field, Duchon $\&$ Robert \cite{Duchon2000Nonlinearity} derived a local form of dissipation which was generalized to incompressible Hall MHD by Galtier \cite{Galtier2018JoPA}. An apparent explanation of anomalous dissipation was given in a work by Cichowlas \textit{et al.} \cite{Cichowlas2005PRL} who showed that even for ideal Eulerian flows, the large-wavenumber modes play the role of an effective eddy viscosity leading to an approximate Kolmogorov scaling at the inertial scales. This idea was further extended to two-dimensional MHD by Krstulovic \textit{et al.} \cite{Krstulovic2011PRE}.
Here we will examine this apparently singular 
behavior (see e.g., \cite{Frisch, Kerr, Dmitruk2005PoF,Mininni2009PRE})
not from a rigorous mathematical perspective,
but from an empirical perspective based on accurate spectral method 
computations.   

Taylor first suggested, based on empirical arguments, that the dissipation rate in a fully developed turbulent system becomes independent of viscosity. K\'arm\'an and Howarth~\cite{Karman1938PRSL} established Taylor's results more rigorously, assuming that the shape of the two-point correlation function remains unchanged. The first experimental verification of K\'arm\'an and Howarth's result came from an experiment performed by Batchelor and Townsend~\cite{Batchelor1953book, Batchelor1947, Batchelor1948a, Batchelor1948b} using wind tunnel measurements. Politano and Pouquet~\cite{Politano1998PRE} generalized the von K\'arm\'an-Howarth equations for isotropic MHD. Wan \textit{et al.}~\cite{Wan2012JFM} investigated nonuniversality of K\'arm\'an-Howarth like decay in MHD in the presence of factors like mean magnetic field, helicity, etc. 

Recently, Wu \textit{et al.} \cite{Wu2013PRL} and Parashar \textit{et al.} \cite{Parashar:ApJL2015} have shown using particle in cell (PIC) simulations that even in the case of weakly collisional plasmas, the von K\'arm\'an decay law remains valid. Most of these studies were interested in the temporal behavior of the fluctuation amplitude and its relationship with the large-scale fluctuations and an energy-containing length scale. For neutral fluids, the variation of dimensionless dissipation rate, $C_{\epsilon}$ with Reynolds number was studied experimentally in shear flows ~\cite{Pearson2002PoF}, numerically in homogeneous incompressible flows~\cite{Kaneda2003PoF}, and in weakly compressible flows~\cite{Pearson2004PoF}. In all cases, an asymptotic value of $C_{\epsilon} \approx 0.5$ was found. We remark here that the dimensionless dissipation rate, $C_{\epsilon}$, as defined in Pearson \textit{et al.}~\cite{Pearson2004PoF} is proportional to one of the constants in von K\'arm\'an decay phenomenology (See Appendix B of ~\cite{Usmanov2014ApJ}). Mininni and Pouquet~\cite{Mininni2009PRE} carried out direct numerical simulations (DNS) of decaying isotropic MHD turbulence, demonstrating that the mean dissipation rate per unit mass, $\epsilon$, remains finite and becomes independent of viscosity and resistivity as the Reynolds number (Taylor-scale Reynolds number, $R_{\lambda}$, in this case) increases. Dallas and Alexakis~\cite{Dallas2014ApJ} performed a similar analysis showing the variation of dimensionless dissipation rate with Taylor-scale Reynolds number with initial velocity and current density being highly correlated. A comparison with \cite{Mininni2009PRE} revealed that the dimensionless dissipation rate saturates to a finite value but the level of saturation depends on the strength of initial cross-correlation.  Recently, Linkmann \textit{et al.}~\cite{Linkmann2015PRL, Linkmann2017PRE} have performed a series of investigations for similar analysis in isotropic MHD. By fitting a model equation, an asymptotic value of dimensionless dissipation rate, $C_{\epsilon, \infty} = 0.265\pm0.013$, was found for nonhelical decaying MHD with no mean field. This value is considerably different from the fluid case. McComb \textit{et al.}~\cite{McComb2017Arxiv} derived a similar model equation for fluid turbulence.

The presence of a mean magnetic field suppresses the decay rate in MHD systems and renders the system anisotropic. 
Early studies by Hossain \textit{et al.}~\cite{Hossain1995PoP, Hossain1996AIP} at low Reynolds number demonstrated that the mean field initially inhibits dissipation, but the effect of Alfv\'enic propagation soon saturates 
for sufficiently large values of the mean field. More recently, Bigot \textit{et al.}~\cite{Bigot2008PRL, Bigot2008PRE, Bigot2011PRE} investigated the energy decay in the presence of a strong uniform magnetic field. Zhdankin \textit{et al.}~\cite{Zhdankin2017MNRAS} also studied the effect of a mean field at 
dissipation sites in MHD turbulence. 
Naturally, one would expect a decrease of the dimensionless dissipation rate 
(or, the von K\'arm\'an decay constant) with increasing strength of mean field.
We present here a further analysis of the dissipation rate for increasing 
values of mean field. 
We find that for each value of the mean field, the dissipation rate asymptotes to a finite value in the limit of large Reynolds number. These results are relevant for systems in which the anisotropy due to presence of a mean magnetic field cannot be neglected. Such situations are often realized in astrophysical plasmas near stellar objects. Our results show that the value of the dimensionless dissipation rate is well separated from the isotropic value for a strong mean field. We first reestablish the isotropic case by comparing with results from \cite{Linkmann2017PRE} and then proceed 
in similar fashion to anisotropic cases.

\section{Equations and Approach}\label{sec:Eqns}

The equations of three dimensional incompressible MHD, in the absence of external forcing, are written as,
\begin{eqnarray} 
\frac{\partial \mathbf{u}}{\partial t} + (\mathbf{u} \cdot \mathbf{\nabla})\mathbf{u} & = &  - \mathbf{\nabla} P + (\mathbf{j} \times \mathbf{B}) + \nu \mathbf{\nabla}^2 \mathbf{u}, \label{eq:u_non}\\ 
\frac{\partial \mathbf{B}}{\partial t} + (\mathbf{u} \cdot \mathbf{\nabla})\mathbf{B} & = & (\mathbf{B} \cdot \mathbf{\nabla})\mathbf{u} + \mu \mathbf{\nabla}^2 \mathbf{B}, \label{eq:induc}\\
\mathbf{\nabla} \cdot \mathbf{u} & = & 0, \label{eq:inc}\\
\mathbf{\nabla} \cdot \mathbf{B} & = & 0 \label{eq:b_gauss},
\end{eqnarray}
where $\mathbf{u}$ is the fluctuating velocity field,  $\mathbf{B}$ is the total magnetic field which we assume can be decomposed into a spatially
uniform mean and a fluctuating part, 
$\mathbf{B}(\mathbf{r},t)=\mathbf{B_0}+\mathbf{b}(\mathbf{r},t)$. Without any loss of generality, we choose $\mathbf{B_0}=B_0\mathbf{ \hat{z}}$. $P$ is the total (thermal+magnetic) pressure field, $\mathbf{j}=(\mathbf{\nabla} \times \mathbf{B})/\mu_0$ is the current density, $\nu$ is the kinematic viscosity, and $\mu$ is the magnetic diffusivity. Eq. (\ref{eq:inc}) enforces incompressibility. For simplicity, we assume the fluid density $\rho$ is constant and set it to unity. For this study, we consider unit magnetic Prandtl number $(\mathrm{Pm} = \nu/\mu)$, i.e., equal viscosity and resistivity,
\begin{eqnarray}
	\nu = \mu \label{eq:pm_1}.
\end{eqnarray}
With this constraint, the MHD equations Eqs.~(\ref{eq:u_non})--(\ref{eq:b_gauss}) can be written in terms of the Elsasser variables, $\mathbf{z_{\pm}}=\mathbf{u}\pm\mathbf{b}$, as,
\begin{eqnarray}
\frac{\partial \mathbf{z}_{\pm}}{\partial t} =  \pm \mathbf{v_A} \cdot \mathbf{\nabla} \mathbf{z}_{\pm} - \mathbf{z}_{\mp} \cdot \mathbf{\nabla} \mathbf{z}_{\pm} - \mathbf{\nabla} P +	 \nu \mathbf{\nabla}^2 \mathbf{z}_{\pm}.\label{eq:mhd_zpm}
\end{eqnarray}
Here, $\mathbf{v_{A}}$ is the Alfv\'en velocity defined as 
$\mathbf{v_{A}} = \mathbf{B_0}/\sqrt{4\pi}$. The Elsasser energies are given by
\begin{eqnarray}
Z^2_{\pm} = \langle |\mathbf{z}_{\pm}|^2 \rangle = \langle |\mathbf{u}\pm\mathbf{b}|^2 \rangle. \label{eq:energy_zpm}
\end{eqnarray}
Following the argument associated 
with preservation of the functional form of the two-point correlation functions, one can generalize the von K\'arm\'an-Howarth~\cite{Karman1938PRSL} result to isotropic MHD~\cite{Politano1998PRE} or MHD isotropic in a plane perpendicular to a strong mean field~\cite{Wan2012JFM},
\begin{eqnarray}
	\frac{d Z^2_{\pm}}{d t} &=& -\alpha_{\pm} \frac{Z^2_{\pm} Z_{\mp}}{L_{\pm}}\label{eq:vk1},\\
	\frac{d L_{\pm}}{d t} &=& \beta_{\pm} Z_{\mp}\label{eq:vk2},
\end{eqnarray}
where $\alpha_{\pm}$ and $\beta_{\pm}$ are constants, and $L_{\pm}$ are the energy-containing scales corresponding to the two Elsasser variables. Precise definitions of $L_{\pm}$ depend on the phenomenology used. We shall come back to the definition of energy-containing scale later in the paper. For small cross helicity $(H_c \simeq 0)$, the $``+"$ and $``-"$ variables remain almost equal, so that one can write,
\begin{eqnarray}
\frac{d Z^2}{d t} &=& -\alpha \frac{Z^3}{L}\label{eq:z_evolve},\\
\frac{d L}{d t} &=& \beta Z\label{eq:L_evolve},
\end{eqnarray}
 where, $Z \simeq Z_{+} \simeq Z_{-}$. 
It is apparent from these equations that the sequence of assumptions
(especially, self-preservation of the correlations) 
leading to this point 
implies that the energy dissipation rate becomes independent of the 
dissipation coefficients $\nu = \mu$.
It is one of the main goals of this paper to measure the finite energy dissipation rate in the 
asymptotic limit of large Reynolds number for MHD in presence of a mean field at zero magnetic and cross helicity.

The mean turbulent energy dissipation rate per unit mass in 
incompressible MHD, 
for unit magnetic Prandtl number $(\mathrm{Pm})$, can be written as,
\begin{eqnarray}
	\epsilon (t) = \nu \langle \omega^2 + j^2 \rangle \label{eq:dissipation},
\end{eqnarray}
where $\mathbf{\omega}=\mathbf{\nabla} \times \mathbf{u}$ is the vorticity and $\langle \cdots \rangle$ denotes spatial averaging.

Linkmann \textit{et al.}~\cite{Linkmann2015PRL, Linkmann2017PRE} proposed a definition of the non-dimensional energy dissipation rate in MHD turbulence as,
\begin{eqnarray}
C_{\epsilon}(t) &=& C^{+}_{\epsilon}(t) + C^{-}_{\epsilon}(t),\label{eq:Ceps_gen}
\end{eqnarray}
where
\begin{eqnarray}
C^{\pm}_{\epsilon}(t) &=& \frac{ \epsilon(t) L_{\pm}(t)}{W_{\pm}(t)^2 W_{\mp}(t)},\label{eq:Ceps_pm}
\end{eqnarray}
where 
\begin{eqnarray}
L_{\pm}(t) = \frac{3\pi}{8 E_{\pm}(t)} \int k^{-1} \langle|\mathbf{z}_{\pm}(\mathbf{k},t)|^2\rangle d\mathbf{k} \label{eq:Lpm},
\end{eqnarray}
where $\mathbf{z}_{\pm}(\mathbf{k},t)$ denotes the Fourier transform of the respective Elsasser variables. Assuming isotropy, the rms values of the Cartesian \emph{components} of $Z_{\pm}^2$ are given by, 
\begin{eqnarray}
W^2_{\pm} = Z^2_{\pm}/3 \label{eq:wpm}. 
\end{eqnarray}
The $W_{\pm}$ are the Elsasser variable analogs of the $u^{\prime}$ in $E^{u} = 3(u^{\prime})^{2}/2$, which is the classical definition often employed in isotropic hydrodynamic work~\cite{Batchelor1953book}. Here we use
\begin{eqnarray}
E_{\pm}(t) = 2 E(t) \pm 2 H_c(t) \equiv Z^2_{\pm} \label{eq:Epm},
\end{eqnarray}
where $E(t)$ is the total energy
\begin{eqnarray}
E(t) = \frac{1}{4} \int \langle|\mathbf{z}_{+}(\mathbf{k},t)|^2 + |\mathbf{z}_{-}(\mathbf{k},t)|^2\rangle d\mathbf{k} \label{eq:Etot},
\end{eqnarray}
and $H_c(t)$ is the cross helicity
\begin{eqnarray}
H_c(t) = \frac{1}{4} \int \langle|\mathbf{z}_{+}(\mathbf{k},t)|^2 - |\mathbf{z}_{-}(\mathbf{k},t)|^2\rangle d\mathbf{k} \label{eq:Hc}.
\end{eqnarray}
For $H_c \simeq 0$, one expects $C^{+}_{\epsilon} \simeq C^{-}_{\epsilon}$. Further, Linkmann \textit{et al.}~\cite{Linkmann2015PRL, Linkmann2017PRE} also suggested a generalized Reynolds number defined as
\begin{eqnarray}
R_{-} &=& \frac{W_{-} L_{+}}{(\nu+\mu)}\label{eq:Re_minus},
\end{eqnarray}
In this paper we have studied the variation of dimensionless dissipation rate, as defined by Eqs.~(\ref{eq:Ceps_gen})--(\ref{eq:Ceps_pm}) with generalized Reynolds number defined in Eq.~(\ref{eq:Re_minus}). For comparison, we also report the integral scale Reynolds number  defined as
\begin{eqnarray}
R_{L} &=& \frac{u^{\prime} L_{\mathrm{int}}}{\nu}\label{eq:Re_int},
\end{eqnarray}
where
\begin{eqnarray}
L_{\mathrm{int}} = \bigg(\frac{3\pi}{4}\bigg) \frac{\int k^{-1} E^u(k) dk}{\int E^u(k) dk} \label{eq:Lint}.
\end{eqnarray}
and $u^{\prime}$ denotes the rms speed with $E^u = 3(u^{\prime})^2/2 $. Here $E^u(k)$ denotes the omnidirectional kinetic energy spectrum and $E^u = \int E^u(k) dk$ is the total kinetic energy. Similarly, the Taylor microscale Reynolds number is defined as
\begin{eqnarray}
R_{\lambda}  &=& \frac{u^{\prime} \lambda}{\nu}\label{eq:Re_tayl},
\end{eqnarray}
where
\begin{eqnarray}
\lambda =  \bigg[ \frac{5 \int E^{u}(k) dk}{\int k^{2} E^{u}(k) dk}\bigg]^{1/2} = u^{\prime} \sqrt{\frac{15 \nu}{\epsilon}} \label{eq:Ltaylor}.
\end{eqnarray}

The effect of a mean magnetic field on magnetohydrodynamic turbulence was first investigated and quantified by Shebalin \textit{et al.}~\cite{Shebalin1983JPP} in two dimensions, and generalized to three dimensions by Oughton \textit{et al.}~\cite{Oughton1994JFM}. Spectral anisotropy in three dimensions is quantified by defining Shebalin angles, $\theta_{Q}$ as
\begin{eqnarray}
\tan^2 \theta_{Q}(t) = \frac{\sum k^2_{\perp} |\mathbf{Q}(\mathbf{k},t)|^2}{\sum k^2_{\parallel} |\mathbf{Q}(\mathbf{k},t)|^2},
\end{eqnarray} 
where $k_{\parallel}$ is the wave vector component along the direction of the mean magnetic field, $k_{\perp}$ is the wave vector in the plane perpendicular to the mean field, and the summations extend over all values of wave vector $\mathbf{k}$. In the present notation, $k_{\parallel} = k_{z}$ and $k^2_{\perp} = k^2_{x} + k^2_{y}$. In general $\mathbf{Q}$ can be any vector field, although here we consider only velocity field $\mathbf{u}$, and magnetic field $\mathbf{b}$.

We define 
\begin{eqnarray}
r = \frac{2 E^b}{B^2_{0}}\label{eq:r},
\end{eqnarray}
where $E^b = \delta b^2/2$ is the total magnetic field fluctuation energy. Therefore, $\sqrt{r}$ is the ratio of fluctuation amplitude and mean magnetic field strength. 
\begin{eqnarray}
r = \bigg(\frac{\delta b}{B_0}\bigg)^2\label{eq:rsqrt}.
\end{eqnarray}

\section{Numerical Method}\label{sec:Method} 

We solve the MHD equations in a periodic box of dimension $(2\pi)^3$ using a pseudo-spectral method in a Fourier basis. We employ second-order Runge-Kutta (RK2) scheme for time advancement and the $2/3$ rule for dealiasing. Instead of using an automatically adjustable time step, in this study the time step was held constant for a set of runs, but halved when instabilities occurred. All the runs have been initialized with a \textit{modal} spectrum proportional to $ 1/[1+(k/k_{0})^{11/3}]$, where the ``knee'' of the spectrum is $k_{0}=4$, and  only Fourier modes within the band $1 \le k \le 15$ are excited. The total kinetic and magnetic energy are both normalized to 0.5 initially. All the measurements have been made at the time of highest dissipation. In all runs, the kinetic, magnetic, and cross helicities are small initially and remain so during the simulation times considered. Further, since velocity and magnetic fields are generated independently, we may assume that 
cross correlations involving higher derivatives~\cite{Dallas2014ApJ}
are suitably small throughout. 

In all runs, the maximum resolved wavenumber, $k_{\mathrm{max}}$, is greater than the dissipation wavenumber (reciprocal of the Kolmogorov length scale $\eta$) defined as
\begin{eqnarray}
k_{\mathrm{diss}} = \frac{1}{\eta} =\bigg( \frac{\epsilon}{\nu^3}\bigg)^{1/4} \label{eq:kdiss}.
\end{eqnarray}
Wan \textit{et al.}~\cite{Wan2010bPoP} showed that the ratio $k_{\mathrm{max}}/k_{\mathrm{diss}} = k_{\mathrm{max}} \eta$ should be at least three for sufficient numerical accuracy of fourth-order moments. However, for studies of the present kind, as was earlier reported by Pearson \textit{et al.}~\cite{Pearson2004PoF}, we find that $k_{\mathrm{max}} \eta \ge 1$ suffices and increasing resolution further does not substantially 
change the results.
Recall that the more strict requirement proposed by Wan \textit{et al}~\cite{Wan2010bPoP}
pertains mainly to higher-order statistics and coherent structures, 
while accurate portrayal of 
lower-order quantities such as energy spectra, 
which control instantaneous dissipation, 
have less stringent requirements. 
This can also be seen from Table II of ~\cite{Wan2010bPoP} where $k_{\mathrm{diss}}$ becomes  accurate in the range $1 \le k_{\mathrm{max}} \eta \le 1.9$. Since $k^4_{\mathrm{diss}} = \epsilon/ \nu^3$, this implies that dissipation rate also becomes accurate in this range.

The details of the simulations used in 
this study are found in Table~\ref{tab:num_details}. 
\begin{table*}
	\centering
	\caption{Parameters for three dimensional spectral simulations: Mean magnetic field strength $B_0$, grid resolution $N^3$, viscosity and resistivity $\nu$ and $\mu$ (set equal), Taylor scale Reyolds number $R_{\lambda}$, Integral scale Reynolds number $R_{L}$, Generalized Reynolds number $R_{-}$, Dimensionless dissipation rate $C_{\epsilon}$, Shebalin angles for velocity field $\theta_u$, and for magnetic field $\theta_b$, Square of ratio of fluctuation amplitude and mean magnetic field 
$r = (\delta b/B_{0})^2$ (Not Applicable for $B_{0}=0$ case), time step $dt$, Ratio of maximum resolved wavenumber and dissipation wavenumber $k_{\mathrm{max}}\eta$. See text for definitions of the quantities. All measurements have been made near the time of maximum dissipation.}
	\label{tab:num_details}
	\noindent  
	\begin{tabularx}{\linewidth}{XXXXXXXXXXXr}
		\hline \hline
		 $B_0$ & $N^3$ & $\nu=\mu$ & $R_{\lambda}$ & $R_{L}$ & $R_{-}$ & $C_{\epsilon}$ &$\theta_{u}$ &$\theta_{b}$ &$r$ & $dt$  & $k_{\mathrm{max}}\eta$ \\
		   &  & $(\times10^{-3})$ &  & & &  &  &  & & $(\times10^{-4})$ &  \\
		\hline 
		 $0$ & $256^3$ & $2.0$ & $32.53$ & $116.09$ & $89.05$ & $0.456$ &54.7 &54.8 & NA\footnote[1]{} & $10.0$ & $1.83$\\
		 $0$ & $512^3$ & $1.0$ & $48.07$ & $216.26$ & $166.82$ & $0.382$ &54.7 &54.6 & NA & $5.0$ & $2.24$\\ 
		 $0$ & $512^3$ & $0.5$ & $68.44$ & $394.21$ & $308.21$ & $0.330$ &54.8 &54.7 & NA & $5.0$ & $1.36$\\
		 $0$ & $768^3$ & $0.4$ & $76.06$ & $483.43$ & $382.76$ & $0.316$ &54.6 &54.4 &NA & $4.0$ & $1.73$\\
		 $0$ & $768^3$ & $0.3$ & $87.68$ & $622.84$ & $496.21$ & $0.300$ &54.6 &54.5 & NA & $4.0$ & $1.40$\\ 
		 $0$ & $1024^3$ & $0.2$ & $105.7$ & $886.04$ & $713.61$ & $0.285$ &54.9 &55.0 &NA & $2.5$ & $1.39$\\
		 $0$ & $1536^3$ & $0.15$ & $121.68$ & $1126.96$ & $910.97$ & $0.270$ &54.7 &54.6 & NA & $1.5$ & $1.69$\\ 
		 $0$ & $2048^3$ & $0.1$ & $145.63$ & $1605.56$ & $1313.66$ & $0.261$ &54.6 &54.7 &NA & $1.25$ & $1.66$\\
		\hline
		 $1$ & $256^3$ & $2.0$ & $40.59$ & $121.5$ & $88.59$ & $0.356$ &60.2 &61.8 &0.7550 & $10$ & $1.92$\\
		 $1$ & $512^3$ & $1.0$ & $59.94$ & $228.62$ & $168.50$ & $0.298$ &62.8 &63.9 &0.7377 & $5.0$ & $2.36$\\
		 $1$ & $512^3$ & $0.5$ & $86.22$ & $413.14$ & $303.21$ & $0.264$ &64.9 &65.7 &0.7209 & $5.0$ & $1.43$\\
		 $1$ & $768^3$ & $0.4$ & $96.17$ & $506.29$ & $374.40$ & $0.254$ &58.0 &59.9 &0.7138 & $4.0$ & $1.82$\\
		 $1$ & $768^3$ & $0.3$ & $110.80$ & $656.66$ & $485.14$ & $0.249$ &66.1 &66.8 &0.7074 & $4.0$ & $1.47$\\
		 $1$ & $1024^3$ & $0.2$ & $135.28$ & $931.42$ & $688.66$ & $0.236$ &66.6 &67.1 &0.7029 & $2.5$ & $1.45$\\
		 $1$ & $1536^3$ & $0.15$ & $159.67$ & $1273.54$ & $932.60$ & $0.238$ &67.3 &67.6 &0.7060 & $2.0$ & $1.77$\\ 
		 $1$ & $2048^3$ & $0.1$ & $190.99$ & $1811.17$ & $1331.78$ & $0.234$ &67.0 &67.3 &0.6986 & $1.25$ & $1.73$\\ 
		\hline 
		 $3$ & $256^3$ & $2.0$ & $59.15$ & $137.51$ & $96.14$ & $0.216$ &64.6 &65.3 &0.0953 & $2.5$ & $2.18$\\
		 $3$ & $512^3$ & $1.0$ & $90.3$ & $256.40$ & $183.8$ & $0.160$ &69.8 &70.3 &0.0899 & $1.0$ & $2.77$\\
		 $3$ & $512^3$ & $0.5$ & $134.42$ & $463.69$ & $331.23$ & $0.132$ &74.0 &74.5 &0.0870 & $1.0$ & $1.72$\\
		 $3$ & $512^3$ & $0.4$ & $157.05$ & $572.75$ & $408.17$ & $0.120$ &75.1 &75.5 &0.0882 & $1.0$ & $1.48$\\
		 $3$ & $768^3$ & $0.3$ & $177.78$ & $738.42$ & $524.07$ & $0.122$ &76.7 &77.2 &0.0871 & $9.0$ & $1.77$\\
		 $3$ & $1024^3$ & $0.2$ & $216.76$ & $1054.46$ & $752.93$ & $0.116$ &78.2 &78.6 &0.0846 & $0.8$ & $1.77$\\
		 $3$ & $1024^3$ & $0.15$ & $247.75$ & $1367.76$ & $977.78$ & $0.114$ &79.1 &79.4 &0.0839 & $0.8$ & $1.42$\\  
		 $3$ & $1536^3$ & $0.1$ & $299.74$ & $1961.17$ & $1406.34$ & $0.111$ &80.0 &80.3 &0.0838 & $0.5$ & $1.57$\\  
		\hline  
		 $8$ & $256^3$ & $2.0$ & $67.45$ & $141.86$ & $99.52$ & $0.169$ &63.2 &63.5 &0.013 & $1.0$ & $2.35$\\
		 $8$ & $512^3$ & $1.0$ & $98.98$ & $247.13$ & $176.00$ & $0.131$ &69.3 &69.7 &0.0125 & $0.5$ & $2.96$\\
		 $8$ & $512^3$ & $0.5$ & $159.41$ & $459.39$ & $327.15$ & $0.094$ &73.8 &74.0 &0.0124 & $0.5$ & $1.86$\\
		 $8$ & $512^3$ & $0.4$ & $190.30$ & $572.86$ & $407.77$ & $0.082$ &74.8 &75.0 &0.0126 & $0.5$ & $1.61$\\
		 $8$ & $512^3$ & $0.3$ & $225.92$ & $725.48$ & $514.83$ & $0.074$ &76.6 &76.7 &0.0124 & $0.5$ & $1.33$\\
		 $8$ & $768^3$ & $0.2$ & $284.55$ & $1014.84$ & $726.54$ & $0.064$ &78.3 &78.4 &0.0123 & $0.3$ & $1.49$\\
		\hline
		 $15$ & $256^3$ & $2.0$ & $66.5$ & $138.08$ & $96.15$ & $0.173$ &63.7 &64.0 &0.0037 & $0.65$ & $2.34$\\
		 $15$ & $512^3$ & $1.0$ & $101.34$ & $249.22$ & $178.09$ & $0.125$ &69.6 &69.8 &0.0036 & $0.3$ & $2.99$\\
		 $15$ & $512^3$ & $0.5$ & $163.05$ & $461.06$ & $327.82$ & $0.090$ &73.9 &74.3 &0.0036 & $0.3$ & $1.88$\\
		 $15$ & $512^3$ & $0.4$ & $184.57$ & $556.69$ & $397.84$ & $0.084$ &74.8 &74.9 &0.0035 & $0.3$ & $1.61$\\
		 $15$ & $512^3$ & $0.3$ & $228.77$ & $727.64$ & $519.81$ & $0.072$ &76.4 &76.6 &0.0035 & $0.3$ & $1.33$\\
		 $15$ & $768^3$ & $0.2$ & $292.14$ & $1023.87$ & $730.58$ & $0.062$ &78.4 &78.4 &0.0035 & $0.25$ & $1.51$\\
		\hline \hline
	\end{tabularx}
\footnotetext{NA $\equiv$  Not Applicable}
\end{table*}
Measurements are made at instant of peak dissipation, and shortly after this time the 
simulations are stopped. 
Ideally, one should consider performing an ensemble of runs and then calculate the statistics. However, the computational cost being prohibitively expensive, particularly for the high mean field cases, we defer such refinements to 
a later time.  

\section{Results}\label{sec:Results} 

\begin{figure}
	\includegraphics[scale=1.0]{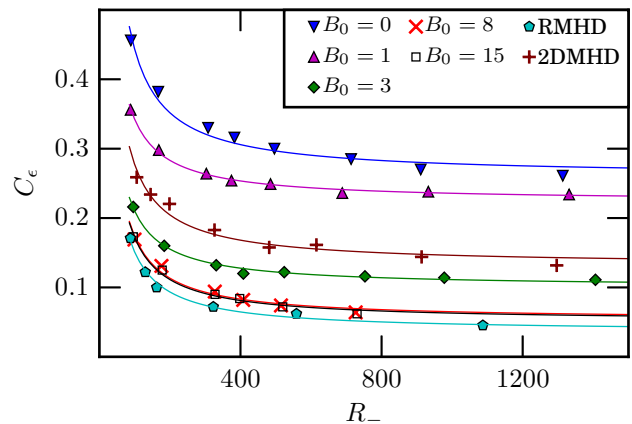}
	\caption{Dimensionless dissipation rate $(C_{\epsilon})$ plotted as a function of generalized Reynolds number $R_{-}$ for three-dimensional runs with different values of mean field $B_0$, one set of two-dimensional runs, and one set of reduced MHD runs  along with fitted polynomials represented by continuous lines of same color as the corresponding runs.}
	\label{fig:ceps}
\end{figure}

The dimensionless dissipation rates, obtained from the runs in Table~\ref{tab:num_details}, 
are plotted as function of the generalized Reynolds number in 
Fig.~\ref{fig:ceps}.
Linkmann \textit{et al.}~\cite{Linkmann2015PRL, Linkmann2017PRE} showed that for isotropic MHD one can fit a simple model equation to $C_{\epsilon}$, provided $R_{-} \geq 80$,
\begin{eqnarray}
	C_{\epsilon} = C_{\epsilon, \infty} + \frac{A}{R_{-}} + O(R^{-2}_{-})\label{eq:Ceps_model},
\end{eqnarray}
where $C_{\epsilon, \infty}$ is the asymptotic value of $C_{\epsilon}$ as $R_{-}$ tends to $\infty$, $A$ is a time dependent coefficient, and $O(R^{-2}_{-})$ represents terms of higher order, which we are neglecting here. 
For lower Reynolds numbers, second order and higher terms cannot be neglected and may be important to derive model equations such as those described in ~\cite{Linkmann2015PRL, Linkmann2017PRE}. Since the main goal of this paper is not to build such model equations, but to measure the asymptotic value of the dimensionless dissipation rate and investigate its variation with mean magnetic field, we consider only high Reynolds numbers $(R_{-} \ge 80)$ and work only with the leading-order term.

We fit Eq.~(\ref{eq:Ceps_model}) to each set of runs; see Table~\ref{tab:num_details}. 
The solid lines in Fig.~\ref{fig:ceps} are obtained from this fitting. We use a least-square fitting technique to fit the polynomial Eq.~(\ref{eq:Ceps_model}) to the data for each value of the mean magnetic field, which ranges over values $0, 1, 3, 8,$ and $15$. 
In all cases we see that the fits agree reasonably well with
the the sets of individual data points, and in some cases the agreement is excellent. 
The asymptotic values, i.e.,
the values of $C_{\epsilon, \infty}$, obtained by this procedure, are reported in Table~\ref{tab:Ceps}.
\begin{table}
	\centering  
	\caption{Summary of the asymptotic non-dimensional dissipation rate $C_{\epsilon, \infty}$ from different simulation sets of Table~\ref{tab:num_details}, and sets of RMHD and 2DMHD runs. Additionally, we report the value of the constant $A$ in Eq.~(\ref{eq:Ceps_model}).}
	\noindent
	\begin{tabularx}{\linewidth}{X X X}
		\hline \hline
		$B_0$ & $C_{\epsilon,\infty}$ & $A$ \\
		\hline
		$0.0$ & $0.260 \pm 0.007$ & $18 \pm 1$\\
		$1.0$ & $0.224 \pm 0.002$ & $11.8 \pm 0.4$\\
		$3.0$ & $0.100 \pm 0.002$ & $11.0 \pm 0.5$\\
		$8.0$ & $0.053 \pm 0.006$ & $12 \pm 1$\\
		$15.0$ & $0.051 \pm 0.005$ & $12.1 \pm 0.9$\\
		RMHD & $0.036 \pm 0.005$ & $11.5 \pm 0.8$\\
		2DMHD & $0.132 \pm 0.007$ & $14 \pm 1$\\
		\hline \hline
	\end{tabularx}
	\label{tab:Ceps}
\end{table}
The value for the isotropic case is in agreement with the one reported in \cite{Linkmann2017PRE}, within  uncertainty. The  values of $C_{\epsilon, \infty}$ for $B_{0} > 0$ decrease first with increasing strength of the mean field, but then saturate. This can be seen by comparing the two sets of data obtained for $B_{0} = 8.0$ and $B_{0}=15.0$, which almost lie on top of each other. Quantitatively, it can be seen from Table~\ref{tab:Ceps} that the two values of $C_{\epsilon}$ for the $B_{0}=8.0$ and $B_{0}=15.0$ case lie within their limits of simulation uncertainty. This result is reminiscent of the Hossain \textit{et al.}~\cite{Hossain1996AIP} study, that observed a suppression of dissipation rate due to a mean field of moderate strength compared to the dissipation rate in the isotropic case. This effect, however, saturates at higher mean field strengths. This is explained by noticing that the mean magnetic field suppresses spectral transfer parallel to the mean field, so that the spectrum becomes progressively more anisotropic, saturating at an anisotropy determined by the parallel bandwidth of the initial data (or forcing) \cite{Shebalin1983JPP,Oughton1994JFM,Bigot2011PRE}. Once saturated, further increase of the mean field has negligible effect. Thus, at strong mean field values, the Alfv\'en time does not play a dominant role in determining the triple decorrelation time that establishes the spectral transfer rates (see e.g., \cite{Zhou2004RMP}).
\begin{figure*}
	\includegraphics[scale=0.9]{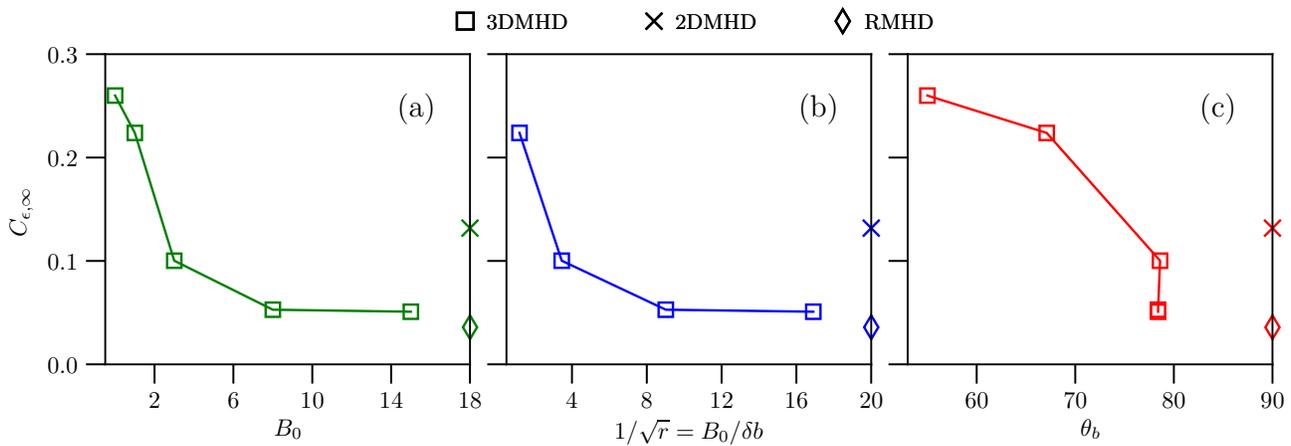}
	\caption{Asymptotic value of the dimensionless dissipation rate $C_{\epsilon}$ plotted against different measures of anisotropy. Panel (a): $C_{\epsilon, \infty}$ versus mean magnetic field strength $B_0$, panel (b): $C_{\epsilon, \infty}$ versus  $1/\sqrt{r}=B_0/\delta b$, panel (c): $C_{\epsilon, \infty}$ versus Shebalin angles $\theta_{b}$. The square, cross, and diamond symbols represent results from the 3DMHD, RMHD, and 2DMHD runs. The RMHD and 2DMHD values are plotted at the edge of the horizontal axes.}
	\label{fig:ceps_r_log}
\end{figure*}
We note here that for the simple case of unit Alfv\'en ratio $(r_{\mathrm{A}}=E^{u}/E^{b}=1)$ and zero cross helicity $(H_c = 0)$, as considered here, similar convergence is obtained if the traditional, hydrodynamic, definition of dimensionless dissipation rate, $C_{\epsilon}=\epsilon {u^{\prime}}^{3}/L_{\mathrm{int}}$, is used instead of Eq.~(\ref{eq:Ceps_gen})--(\ref{eq:Ceps_pm}). However, this may not be necessarily true for any general condition.

To compare the convergence of the non-dimensional dissipation rate in three-dimensional MHD in the presence of a strong mean field with two-dimensional MHD (2DMHD), we perform a set of 2DMHD simulations with decreasing viscosity and resistivity (See \cite{Shebalin1983JPP} for governing equations and other details). The two-dimensional simulations are performed in almost identical conditions as the three dimensional ones. Here, the box dimension is $(2\pi)^2$ and the initial spectrum is excited in $1 \le k \le 15$ band, with the \textit{modal} spectrum proportional to $ 1/[1+(k/4)^{8/3}]$ so that the omni-directional spectrum is $\sim k^{-5/3}$ at large $k$. As can be seen in Fig.~\ref{fig:ceps}, the values corresponding to 2DMHD are close to neither the $B_0 = 0$ case nor the $B_0 = 8,15$ limit. This result demonstrates the unique nature of two-dimensional decay of magnetofluids. Although spectral transfer becomes progressively two-dimensional as the mean magnetic field is increased, the limit is not identical with the perfectly two-dimensional system. This discontinuity is presumably due to the additional invariant in 2DMHD, namely the mean squared magnetic potential which is not conserved for 3DMHD, even in the presence of a strong mean field. This extra invariant puts additional constraints on the dynamics of decay of energy-containing eddies in 2DMHD. In particular, with regard to the presence of a second ideal invariant in 2D, we note that the mean-square potential is expected to be back-transfered to longer wavelengths during the cascade. This induces the growth of large magnetic islands, which contribute at most very weakly to the direct cascade. Therefore, one might view the standard estimate of the similarity scale (i.e., energy-containing scale),  given in Eq.~(\ref{eq:Lpm}), as somewhat of an overestimate. This may point the way to understanding the overestimate given by Eq.~(\ref{eq:Lpm}) when applied to the 2D case.

To investigate the issue of reduction of dimensionality further, we also employ a set of reduced MHD (RMHD) simulations. RMHD is often used as a simpler model of plasma in the presence of an energetically strong mean magnetic field (see \cite{Oughton2017ApJ} for governing equations and other discussions). Again, the RMHD runs have almost identical conditions as the 3DMHD ones except that here, the measurements have been made near the instant of maximum dissipation associated with the $k_{\parallel}=1$ modes. Here, we set $B_0 = 1$, noting that this is in rescaled ``code'' units. In unscaled units this corresponds to physical situations with $ \delta b  \ll B_0 $, as must be the case for RMHD \cite[e.g.,][]{Oughton2017ApJ}. Fig.~\ref{fig:ceps} shows that, contrary to the 2DMHD results, the RMHD values almost superimpose on the 3DMHD values for high mean field cases. These results show that, at least for the problem of global dissipation, RMHD may be a more realistic approximation to the full three-dimensional system in the presence of a strong mean magnetic field than 2DMHD is.

An additional view of the approach to asymptotic values of the 
dissipation rate is provided in Fig.~\ref{fig:ceps_r_log}. Panel (a) shows the empirical estimate of the asymptotic decay rate  
$C_{\epsilon,\infty}$ plotted against the mean field strength $B_0$, and panel (b) shows $C_{\epsilon,\infty}$ plotted versus $1/\sqrt{r} = B_0/\delta b$.
Still another view 
is provided by examining how the asymptotic decay rate varies 
with spectral anisotropy measured by the value of the 
Shebalin angle $\theta_b$ at the time of peak dissipation; 
this is  shown in 
Figure \ref{fig:ceps_r_log}, panel (c).
Here we see that the anisotropy saturates at large mean field values,
so that the subsequent values of dissipation rate cluster near the associated 
saturated Shebalin angle. Since the values of the Shebalin angles also loosely depend on the Reynolds number, the values of $\theta_b$ corresponding to $\nu=0.0002$ are considered here.

In Fig.~\ref{fig:ceps_r_log} we also show the values of $C_{\epsilon,\infty}$, extrapolated from the 2DMHD and RMHD runs. The values of $\delta b/B_0$ and the ratio of perpendicular to parallel characteristic length scales $\ell_{\perp}/\ell_{\parallel}$ are small but somewhat arbitrary for RMHD and there are an infinite choice of values. The 2DMHD values are only for comparison. Therefore, we have plotted the $C_{\epsilon,\infty}$ corresponding to the 2DMHD and RMHD simulations at the right edge of the horizontal axes of the three subplots, arbitrarily.

\section{Summary}\label{sec:Summary}

We have performed a quantitative measurement of variation of the non-dimensional dissipation rate in MHD turbulence in presence of a mean magnetic field using Fourier pseudo-spectral simulations. 
We find 
that the dissipation
rate approaches a nonzero asymptotic value for increasing 
Reynolds numbers (mechanical and magnetic), and for 
increasing mean DC magnetic field strength. This 
confirms the generalizations of the von K\'arm\'an-Howarth 
theory of hydrodynamics to the case of magnetohydrodynamics~\cite{Hossain1995PoP,Hossain1996AIP}. 

This conclusion 
provides essential confirmation of 
the underlying theory, normally assumed, 
upon which research is based
in several key areas of space and astrophysics.
Two prominent examples are the 
derivation and use of so-called ``third-order laws'' 
or Yaglom relations, such as the well studied Politano-Pouquet relations
\cite{Politano1998PRE}, 
and the estimation of decay rates in turbulence transport 
theories~\cite{Breech2008JGR}. 
Generalized Yaglom laws have been widely employed in 
studies of interplanetary turbulence \cite{MacBride2008ApJ,Sorriso-Valvo2007PRL,Coburn2014PRS},
while turbulence-based global models have found important applications
(as well as agreement with observations) 
in simulation of the outer heliospheric plasma~\cite{Breech2008JGR,vanderHolst2014ApJ,Lionello2014ApJ}, 
and in the inner solar wind and corona~\cite{Usmanov2014ApJ}.

We note that the saturation of dimensionless dissipation rate is obtained for low values of $\delta b/B_{0}$ where weak turbulence could become important in certain circumstances~\cite{Galtier2000JPP,Meyrand2016PRL}. For weak turbulence, the leading order behavior is that of waves, and energy transfer across scales is achieved through resonant interactions among wave modes, mediated by interaction with 2D (non-propagating) modes or quasi-2D modes. Although the present simulation results, including the RMHD cases, fall within the strong-turbulence regime (due to non-propagating fluctuations, see \cite{Dmitruk:PoP2009}), one might expect a similar dissipation anomaly in weak turbulence. However, we have not examined the weak turbulence regime here.

Clearly there is scope for application of the present findings 
to numerous astrophysical plasmas. 
As discussed before, the Reynolds number in these systems are very large and often these systems have a strong mean magnetic field along with the turbulent component. For example, the ratio of the rms fluctuation to the mean magnetic field in the solar wind at 1 AU is $\sim 0.5$. Other astrophysical systems can be even more anisotropic. The solar corona has an even 
stronger mean magnetic field, which makes the plasma highly anisotropic in those regions. The Parker Solar Probe (PSP) Mission, which was launched on August 12, 2018, will make in-situ measurements close to the Sun, in the solar corona. 
The results presented in this paper will be helpful to investigate PSP data and for modeling the solar corona. 

The analysis presented here also 
enables one to directly probe the validity 
of the dissipative anomaly, that is,
the assumption of finite dissipation at 
(approaching) infinite Reynolds number for anisotropic MHD. 
The same may not be true for pure hydrodynamic turbulence~\cite{Dmitruk2005PoF}, since two dimensional hydrodynamic turbulence, unlike its MHD counterpart, does not maintain finite dissipation in the limit of vanishing viscosity. 

Beyond the explanation based on anisotropic spectral transfer given above, the saturation of dissipation rate may also be related to the reconnection rate in the system, especially since magnetic reconnection provides an efficient mechanism of dissipation. We note that as the mean magnetic field becomes stronger, the system becomes quasi-two dimensional and the islands may begin to form thin current sheets, facilitating reconnection in the classical sense. It would be interesting to investigate further how the statistics of the reconnection rates change as the strength of the mean field increases (see \cite{Smith2004GRL}).

It may be 
interesting to perform a study similar to the current one for 
other systems that become anisotropic. 
One example is rotating turbulence. 
It has been shown that under rotation, a turbulent fluid system becomes 
anisotropic and approaches a two dimensional state, similar to anisotropic MHD
\cite{Zhou1995PoF}. 

Another important direction of further investigation is eddy viscosity which is often used in global and homogeneous turbulence simulations as an approximation to represent effects of unresolved scales. In order for the results of an eddy viscosity or any other such models to be physical, it is important that the results presented in this paper are maintained in such approximations, at least qualitatively. We plan to address these problems using Large-Eddy Simulations (LES) in future.

\section*{Acknowledgments}
This research is supported in part by NSF AGS-1063439 and AGS-1156094 (SHINE) and by NASA grants NNX15AB88G
 (LWS), NNX14AI63G (Heliophysics Grand Challenges), NNX17AB79G (Heliophysics GI Program), 
 and the Solar Probe Plus project under subcontract SUB0000165 from the Princeton University IS$\odot$IS project. M.W. is supported by NSFC Grants $11672123$ and $91752201$. We would like to acknowledge high-performance computing support from Cheyenne (doi:10.5065/D6RX99HX) provided by NCAR's Computational and Information Systems Laboratory, sponsored by the National Science Foundation.


\begin{thebibliography}{0}%
\makeatletter
\providecommand \@ifxundefined [1]{%
 \@ifx{#1\undefined}
}%
\providecommand \@ifnum [1]{%
 \ifnum #1\expandafter \@firstoftwo
 \else \expandafter \@secondoftwo
 \fi
}%
\providecommand \@ifx [1]{%
 \ifx #1\expandafter \@firstoftwo
 \else \expandafter \@secondoftwo
 \fi
}%
\providecommand \natexlab [1]{#1}%
\providecommand \enquote  [1]{``#1''}%
\providecommand \bibnamefont  [1]{#1}%
\providecommand \bibfnamefont [1]{#1}%
\providecommand \citenamefont [1]{#1}%
\providecommand \href@noop [0]{\@secondoftwo}%
\providecommand \href [0]{\begingroup \@sanitize@url \@href}%
\providecommand \@href[1]{\@@startlink{#1}\@@href}%
\providecommand \@@href[1]{\endgroup#1\@@endlink}%
\providecommand \@sanitize@url [0]{\catcode `\\12\catcode `\$12\catcode
  `\&12\catcode `\#12\catcode `\^12\catcode `\_12\catcode `\%12\relax}%
\providecommand \@@startlink[1]{}%
\providecommand \@@endlink[0]{}%
\providecommand \url  [0]{\begingroup\@sanitize@url \@url }%
\providecommand \@url [1]{\endgroup\@href {#1}{\urlprefix }}%
\providecommand \urlprefix  [0]{URL }%
\providecommand \Eprint [0]{\href }%
\providecommand \doibase [0]{http://dx.doi.org/}%
\providecommand \selectlanguage [0]{\@gobble}%
\providecommand \bibinfo  [0]{\@secondoftwo}%
\providecommand \bibfield  [0]{\@secondoftwo}%
\providecommand \translation [1]{[#1]}%
\providecommand \BibitemOpen [0]{}%
\providecommand \bibitemStop [0]{}%
\providecommand \bibitemNoStop [0]{.\EOS\space}%
\providecommand \EOS [0]{\spacefactor3000\relax}%
\providecommand \BibitemShut  [1]{\csname bibitem#1\endcsname}%
\let\auto@bib@innerbib\@empty
\end{thebibliography}%


\begin{thebibliography}{56}%
	\makeatletter
	\providecommand \@ifxundefined [1]{%
		\@ifx{#1\undefined}
	}%
	\providecommand \@ifnum [1]{%
		\ifnum #1\expandafter \@firstoftwo
		\else \expandafter \@secondoftwo
		\fi
	}%
	\providecommand \@ifx [1]{%
		\ifx #1\expandafter \@firstoftwo
		\else \expandafter \@secondoftwo
		\fi
	}%
	\providecommand \natexlab [1]{#1}%
	\providecommand \enquote  [1]{``#1''}%
	\providecommand \bibnamefont  [1]{#1}%
	\providecommand \bibfnamefont [1]{#1}%
	\providecommand \citenamefont [1]{#1}%
	\providecommand \href@noop [0]{\@secondoftwo}%
	\providecommand \href [0]{\begingroup \@sanitize@url \@href}%
	\providecommand \@href[1]{\@@startlink{#1}\@@href}%
	\providecommand \@@href[1]{\endgroup#1\@@endlink}%
	\providecommand \@sanitize@url [0]{\catcode `\\12\catcode `\$12\catcode
		`\&12\catcode `\#12\catcode `\^12\catcode `\_12\catcode `\%12\relax}%
	\providecommand \@@startlink[1]{}%
	\providecommand \@@endlink[0]{}%
	\providecommand \url  [0]{\begingroup\@sanitize@url \@url }%
	\providecommand \@url [1]{\endgroup\@href {#1}{\urlprefix }}%
	\providecommand \urlprefix  [0]{URL }%
	\providecommand \Eprint [0]{\href }%
	\providecommand \doibase [0]{http://dx.doi.org/}%
	\providecommand \selectlanguage [0]{\@gobble}%
	\providecommand \bibinfo  [0]{\@secondoftwo}%
	\providecommand \bibfield  [0]{\@secondoftwo}%
	\providecommand \translation [1]{[#1]}%
	\providecommand \BibitemOpen [0]{}%
	\providecommand \bibitemStop [0]{}%
	\providecommand \bibitemNoStop [0]{.\EOS\space}%
	\providecommand \EOS [0]{\spacefactor3000\relax}%
	\providecommand \BibitemShut  [1]{\csname bibitem#1\endcsname}%
	\let\auto@bib@innerbib\@empty
	\bibitem [{\citenamefont {Taylor}(1935)}]{Taylor1935}%
	\BibitemOpen
	\bibfield  {author} {\bibinfo {author} {\bibfnamefont {G.~I.}\ \bibnamefont
			{Taylor}},\ }\bibfield  {title} {\textit {\bibinfo {title} {Statistical
				Theory of Turbulence. {Parts} 1--4},}\ }\href {\doibase
		10.1098/rspa.1935.0158} {\bibfield  {journal} {\bibinfo  {journal} {Proc. R.
				Soc. London Ser. A}\ }\textbf {\bibinfo {volume} {151}},\ \bibinfo {pages}
		{421--444} (\bibinfo {year} {1935})}\BibitemShut {NoStop}%
	\bibitem [{\citenamefont {de~K{\'a}rm{\'a}n}\ and\ \citenamefont
		{Howarth}(1938)}]{Karman1938PRSL}%
	\BibitemOpen
	\bibfield  {author} {\bibinfo {author} {\bibfnamefont {T.}~\bibnamefont
			{de~K{\'a}rm{\'a}n}}\ and\ \bibinfo {author} {\bibfnamefont {L.}~\bibnamefont
			{Howarth}},\ }\bibfield  {title} {\textit {\bibinfo {title} {On the
				Statistical Theory of Isotropic Turbulence},}\ }\href {\doibase
		10.1098/rspa.1938.0013} {\bibfield  {journal} {\bibinfo  {journal} {Proc.
				Roy. Soc. London Ser. A}\ }\textbf {\bibinfo {volume} {164}},\ \bibinfo
		{pages} {192--215} (\bibinfo {year} {1938})}\BibitemShut {NoStop}%
	\bibitem [{\citenamefont {Kolmogorov}(1941{\natexlab{a}})}]{Kolmogorov1941c}%
	\BibitemOpen
	\bibfield  {author} {\bibinfo {author} {\bibfnamefont {A.~N.}\ \bibnamefont
			{Kolmogorov}},\ }\bibfield  {title} {\textit {\bibinfo {title} {Dissipation
				of Energy in the Locally Isotropic Turbulence},}\ }\href {\doibase
		10.1098/rspa.1991.0076} {\bibfield  {journal} {\bibinfo  {journal} {C.R.
				Acad. Sci. U.R.S.S.}\ }\textbf {\bibinfo {volume} {32}},\ \bibinfo {pages}
		{16} (\bibinfo {year} {1941}{\natexlab{a}})},\ \bibinfo {note} {[Reprinted in
		Proc.\ R.\ Soc.\ London, Ser.\ A \textbf{434}, 15--17 (1991)]}\BibitemShut
	{NoStop}%
	\bibitem [{\citenamefont {Donzis}\ \emph {et~al.}(2005)\citenamefont {Donzis},
		\citenamefont {Sreenivasan},\ and\ \citenamefont {Yeung}}]{Donzis2005JFM}%
	\BibitemOpen
	\bibfield  {author} {\bibinfo {author} {\bibfnamefont {D.~A.}\ \bibnamefont
			{Donzis}}, \bibinfo {author} {\bibfnamefont {K.~R.}\ \bibnamefont
			{Sreenivasan}}, \ and\ \bibinfo {author} {\bibfnamefont {P.~K.}\ \bibnamefont
			{Yeung}},\ }\bibfield  {title} {\textit {\bibinfo {title} {Scalar
				Dissipation Rate and Dissipative Anomaly in Isotropic Turbulence},}\ }\href
	{\doibase 10.1017/S0022112005004039} {\bibfield  {journal} {\bibinfo
			{journal} {J. Fluid Mech.}\ }\textbf {\bibinfo {volume} {532}},\
		\bibinfo {pages} {199--216} (\bibinfo {year} {2005})}\BibitemShut {NoStop}%
	\bibitem [{\citenamefont {Kolmogorov}(1941{\natexlab{b}})}]{Kolmogorov1941a}%
	\BibitemOpen
	\bibfield  {author} {\bibinfo {author} {\bibfnamefont {A.~N.}\ \bibnamefont
			{Kolmogorov}},\ }\bibfield  {title} {\textit {\bibinfo {title} {Local
				Structure of Turbulence in an Incompressible Viscous Fluid at Very High
				{Reynolds} Numbers},}\ }\href {\doibase 10.1098/rspa.1991.0075} {\bibfield
		{journal} {\bibinfo  {journal} {Dokl. Akad. Nauk SSSR}\ }\textbf {\bibinfo
			{volume} {30}},\ \bibinfo {pages} {301--305} (\bibinfo {year}
		{1941}{\natexlab{b}})},\ \bibinfo {note} {[Reprinted in Proc.\ R.\ Soc.\
		London, Ser.\ A \textbf{434}, 9--13 (1991)]}\BibitemShut {NoStop}%
	\bibitem [{\citenamefont {Ravelet}\ \emph {et~al.}(2008)\citenamefont
		{Ravelet}, \citenamefont {Chiffaudel},\ and\ \citenamefont
		{Daviaud}}]{Ravelet2008JFM}%
	\BibitemOpen
	\bibfield  {author} {\bibinfo {author} {\bibfnamefont {F.}\ \bibnamefont
			{Ravelet}}, \bibinfo {author} {\bibfnamefont {A.}\ \bibnamefont
			{Chiffaudel}}, \ and\ \bibinfo {author} {\bibfnamefont {F.}\
			\bibnamefont {Daviaud}},\ }\bibfield  {title} {\textit {\bibinfo {title}
			{Supercritical Transition to Turbulence in an Inertially Driven von
				K\'arm\'an Closed Flow},}\ }\href {\doibase 10.1017/S0022112008000712}
	{\bibfield  {journal} {\bibinfo  {journal} {J. Fluid Mech.}\
		}\textbf {\bibinfo {volume} {601}},\ \bibinfo {pages} {339--364} (\bibinfo
		{year} {2008})}\BibitemShut {NoStop}%
	\bibitem [{\citenamefont {Saint-Michel}\ \emph {et~al.}(2014)\citenamefont
		{Saint-Michel}, \citenamefont {Herbert}, \citenamefont {Salort},
		\citenamefont {Baudet}, \citenamefont {Bon~Mardion}, \citenamefont {Bonnay},
		\citenamefont {Bourgoin}, \citenamefont {Castaing}, \citenamefont
		{Chevillard}, \citenamefont {Daviaud}, \citenamefont {Diribarne},
		\citenamefont {Dubrulle}, \citenamefont {Gagne}, \citenamefont {Gibert},
		\citenamefont {Girard}, \citenamefont {Hébral}, \citenamefont {Lehner},\
		and\ \citenamefont {Rousset}}]{SaintMichel2014PoF}%
	\BibitemOpen
	\bibfield  {author} {\bibinfo {author} {\bibfnamefont {B.}~\bibnamefont
			{Saint-Michel}}, \bibinfo {author} {\bibfnamefont {E.}~\bibnamefont
			{Herbert}}, \bibinfo {author} {\bibfnamefont {J.}~\bibnamefont {Salort}},
		\bibinfo {author} {\bibfnamefont {C.}~\bibnamefont {Baudet}}, \bibinfo
		{author} {\bibfnamefont {M.}~\bibnamefont {Bon~Mardion}}, \bibinfo {author}
		{\bibfnamefont {P.}~\bibnamefont {Bonnay}}, \bibinfo {author} {\bibfnamefont
			{M.}~\bibnamefont {Bourgoin}}, \bibinfo {author} {\bibfnamefont
			{B.}~\bibnamefont {Castaing}}, \bibinfo {author} {\bibfnamefont
			{L.}~\bibnamefont {Chevillard}}, \bibinfo {author} {\bibfnamefont
			{F.}~\bibnamefont {Daviaud}}, \bibinfo {author} {\bibfnamefont
			{P.}~\bibnamefont {Diribarne}}, \bibinfo {author} {\bibfnamefont
			{B.}~\bibnamefont {Dubrulle}}, \bibinfo {author} {\bibfnamefont
			{Y.}~\bibnamefont {Gagne}}, \bibinfo {author} {\bibfnamefont
			{M.}~\bibnamefont {Gibert}}, \bibinfo {author} {\bibfnamefont
			{A.}~\bibnamefont {Girard}}, \bibinfo {author} {\bibfnamefont
			{B.}~\bibnamefont {Hébral}}, \bibinfo {author} {\bibfnamefont {Th.}\
			\bibnamefont {Lehner}}, \ and\ \bibinfo {author} {\bibfnamefont
			{B.}~\bibnamefont {Rousset}},\ }\bibfield  {title} {\textit {\bibinfo
			{title} {Probing Quantum and Classical Turbulence Analogy in von K\'arm\'an
				Liquid Helium, Nitrogen, and Water Experiments},}\ }\href {\doibase
		10.1063/1.4904378} {\bibfield  {journal} {\bibinfo  {journal} {Phys.
				Fluids}\ }\textbf {\bibinfo {volume} {26}},\ \bibinfo {pages} {125109}
		(\bibinfo {year} {2014})}\BibitemShut {NoStop}%
	\bibitem [{\citenamefont {Matthaeus}\ \emph {et~al.}(2005)\citenamefont
		{Matthaeus}, \citenamefont {Dasso}, \citenamefont {Weygand}, \citenamefont
		{Milano}, \citenamefont {Smith},\ and\ \citenamefont
		{Kivelson}}]{Matthaeus2005PRL}%
	\BibitemOpen
	\bibfield  {author} {\bibinfo {author} {\bibfnamefont {W.~H.}\ \bibnamefont
			{Matthaeus}}, \bibinfo {author} {\bibfnamefont {S.}~\bibnamefont {Dasso}},
		\bibinfo {author} {\bibfnamefont {J.~M.}\ \bibnamefont {Weygand}}, \bibinfo
		{author} {\bibfnamefont {L.~J.}\ \bibnamefont {Milano}}, \bibinfo {author}
		{\bibfnamefont {C.~W.}\ \bibnamefont {Smith}}, \ and\ \bibinfo {author}
		{\bibfnamefont {M.~G.}\ \bibnamefont {Kivelson}},\ }\bibfield  {title}
	{\textit {\bibinfo {title} {Spatial Correlation of Solar-Wind Turbulence
				from Two-Point Measurements},}\ }\href {\doibase
		10.1103/PhysRevLett.95.231101} {\bibfield  {journal} {\bibinfo  {journal}
			{Phys. Rev. Lett.}\ }\textbf {\bibinfo {volume} {95}},\ \bibinfo {pages}
		{231101} (\bibinfo {year} {2005})}\BibitemShut {NoStop}%
	\bibitem [{\citenamefont {Verma}(1996)}]{Verma1996JGR}%
	\BibitemOpen
	\bibfield  {author} {\bibinfo {author} {\bibfnamefont {M.~K.}\
			\bibnamefont {Verma}},\ }\bibfield  {title} {\textit {\bibinfo {title}
			{Nonclassical Viscosity and Resistivity of the Solar Wind Plasma},}\ }\href
	{\doibase 10.1029/96JA02324} {\bibfield  {journal} {\bibinfo  {journal}
			{J. Geophys. Res.}\ }\textbf {\bibinfo {volume}
			{101}},\ \bibinfo {pages} {27543--27548} (\bibinfo {year}
		{1996})}\BibitemShut {NoStop}%
	\bibitem [{\citenamefont {{Onsager}}(1949)}]{Onsagar1949d}%
	\BibitemOpen
	\bibfield  {author} {\bibinfo {author} {\bibfnamefont {L.}~\bibnamefont
			{{Onsager}}},\ }\bibfield  {title} {\textit {\bibinfo {title} {{Statistical
					Hydrodynamics}},}\ }\href@noop {} {\bibfield  {journal} {\bibinfo  {journal}
			{Il Nuovo Cimento}\ }\textbf {\bibinfo {volume} {6}},\ \bibinfo {pages}
		{279--287} (\bibinfo {year} {1949})}\BibitemShut {NoStop}%
	\bibitem [{\citenamefont {Duchon}\ and\ \citenamefont
		{Robert}(2000)}]{Duchon2000Nonlinearity}%
	\BibitemOpen
	\bibfield  {author} {\bibinfo {author} {\bibfnamefont {J.}\ \bibnamefont
			{Duchon}}\ and\ \bibinfo {author} {\bibfnamefont {R.}\ \bibnamefont
			{Robert}},\ }\bibfield  {title} {\textit {\bibinfo {title} {Inertial Energy
				Dissipation for Weak Solutions of Incompressible Euler and Navier-Stokes
				Equations},}\ }\href {\doibase https://doi.org/10.1088/0951-7715/13/1/312}
	{\bibfield  {journal} {\bibinfo  {journal} {Nonlinearity}\ }\textbf {\bibinfo
			{volume} {13}},\ \bibinfo {pages} {249} (\bibinfo {year} {2000})}\BibitemShut
	{NoStop}%
	\bibitem [{\citenamefont {Galtier}(2018)}]{Galtier2018JoPA}%
	\BibitemOpen
	\bibfield  {author} {\bibinfo {author} {\bibfnamefont {S.}\
			\bibnamefont {Galtier}},\ }\bibfield  {title} {\textit {\bibinfo {title} {On
				the Origin of the Energy Dissipation Anomaly in (Hall)
				Magnetohydrodynamics},}\ }\href {\doibase
		https://doi.org/10.1088/1751-8121/aabbb5} {\bibfield  {journal} {\bibinfo
			{journal} {J. Phys. A: Math. Theor.}\ }\textbf
		{\bibinfo {volume} {51}},\ \bibinfo {pages} {205501} (\bibinfo {year}
		{2018})}\BibitemShut {NoStop}%
	\bibitem [{\citenamefont {Cichowlas}\ \emph {et~al.}(2005)\citenamefont
		{Cichowlas}, \citenamefont {Bona\"{\i}ti}, \citenamefont {Debbasch},\ and\
		\citenamefont {Brachet}}]{Cichowlas2005PRL}%
	\BibitemOpen
	\bibfield  {author} {\bibinfo {author} {\bibfnamefont {C.}\ \bibnamefont
			{Cichowlas}}, \bibinfo {author} {\bibfnamefont {P.}\ \bibnamefont
			{Bona\"{\i}ti}}, \bibinfo {author} {\bibfnamefont {F.}\ \bibnamefont
			{Debbasch}}, \ and\ \bibinfo {author} {\bibfnamefont {M.}\ \bibnamefont
			{Brachet}},\ }\bibfield  {title} {\textit {\bibinfo {title} {Effective
				Dissipation and Turbulence in Spectrally Truncated Euler Flows},}\ }\href
	{\doibase 10.1103/PhysRevLett.95.264502} {\bibfield  {journal} {\bibinfo
			{journal} {Phys. Rev. Lett.}\ }\textbf {\bibinfo {volume} {95}},\ \bibinfo
		{pages} {264502} (\bibinfo {year} {2005})}\BibitemShut {NoStop}%
	\bibitem [{\citenamefont {Krstulovic}\ \emph {et~al.}(2011)\citenamefont
		{Krstulovic}, \citenamefont {Brachet},\ and\ \citenamefont
		{Pouquet}}]{Krstulovic2011PRE}%
	\BibitemOpen
	\bibfield  {author} {\bibinfo {author} {\bibfnamefont {G.}\ \bibnamefont
			{Krstulovic}}, \bibinfo {author} {\bibfnamefont {M.}\ \bibnamefont
			{Brachet}}, \ and\ \bibinfo {author} {\bibfnamefont {A.}\ \bibnamefont
			{Pouquet}},\ }\bibfield  {title} {\textit {\bibinfo {title} {Alfv\'en Waves
				and Ideal Two-Dimensional Galerkin Truncated Magnetohydrodynamics},}\ }\href
	{\doibase 10.1103/PhysRevE.84.016410} {\bibfield  {journal} {\bibinfo
			{journal} {Phys. Rev. E}\ }\textbf {\bibinfo {volume} {84}},\ \bibinfo
		{pages} {016410} (\bibinfo {year} {2011})}\BibitemShut {NoStop}%
	\bibitem [{\citenamefont {Morf}\ \emph {et~al.}(1980)\citenamefont {Morf},
		\citenamefont {Orszag},\ and\ \citenamefont {Frisch}}]{Frisch}%
	\BibitemOpen
	\bibfield  {author} {\bibinfo {author} {\bibfnamefont {R.~H.}\ \bibnamefont
			{Morf}}, \bibinfo {author} {\bibfnamefont {S.~A.}\ \bibnamefont {Orszag}}, \
		and\ \bibinfo {author} {\bibfnamefont {U.}~\bibnamefont {Frisch}},\
	}\bibfield  {title} {\textit {\bibinfo {title} {{Spontaneous Singularity in
				Three-Dimensional, Inviscid, Incompressible Flow}},}\ }\href {\doibase
	10.1103/PhysRevLett.44.572} {\bibfield  {journal} {\bibinfo  {journal}
		{Phys. Rev. Lett.}\ }\textbf {\bibinfo {volume} {44}},\ \bibinfo
	{pages} {572--575} (\bibinfo {year} {1980})}\BibitemShut {NoStop}%
\bibitem [{\citenamefont {Kerr}(1993)}]{Kerr}%
\BibitemOpen
\bibfield  {author} {\bibinfo {author} {\bibfnamefont {R.~M.}\ \bibnamefont
		{Kerr}},\ }\bibfield  {title} {\textit {\bibinfo {title} {{Evidence for a
				Singularity of the Three-Dimensional, Incompressible Euler Equations}},}\
}\href {\doibase 10.1063/1.858849} {\bibfield  {journal} {\bibinfo  {journal}
	{Phys. Fluids}\ }\textbf {\bibinfo {volume} {5}},\ \bibinfo {pages}
{1725--1746} (\bibinfo {year} {1993})}\BibitemShut {NoStop}%
\bibitem [{\citenamefont {Dmitruk}\ and\ \citenamefont
	{Montgomery}(2005)}]{Dmitruk2005PoF}%
\BibitemOpen
\bibfield  {author} {\bibinfo {author} {\bibfnamefont {P.}\ \bibnamefont
		{Dmitruk}}\ and\ \bibinfo {author} {\bibfnamefont {D.~C.}\ \bibnamefont
		{Montgomery}},\ }\bibfield  {title} {\textit {\bibinfo {title} {Numerical
			Study of the Decay of Enstrophy in a Two-Dimensional Navier-Stokes Fluid in
			the Limit of Very Small Viscosities},}\ }\href {\doibase 10.1063/1.1864134}
{\bibfield  {journal} {\bibinfo  {journal} {Phys. Fluids}\ }\textbf {\bibinfo
		{volume} {17}},\ \bibinfo {pages} {035114} (\bibinfo {year}
	{2005})}\BibitemShut {NoStop}%
\bibitem [{\citenamefont {Mininni}\ and\ \citenamefont
	{Pouquet}(2009)}]{Mininni2009PRE}%
\BibitemOpen
\bibfield  {author} {\bibinfo {author} {\bibfnamefont {P.~D.}\ \bibnamefont
		{Mininni}}\ and\ \bibinfo {author} {\bibfnamefont {A.}~\bibnamefont
		{Pouquet}},\ }\bibfield  {title} {\textit {\bibinfo {title} {Finite
			Dissipation and Intermittency in Magnetohydrodynamics},}\ }\href {\doibase
	10.1103/PhysRevE.80.025401} {\bibfield  {journal} {\bibinfo  {journal} {Phys.
			Rev. E}\ }\textbf {\bibinfo {volume} {80}},\ \bibinfo {pages} {025401}
	(\bibinfo {year} {2009})}\BibitemShut {NoStop}%
\bibitem [{\citenamefont {Batchelor}(1953)}]{Batchelor1953book}%
\BibitemOpen
\bibfield  {author} {\bibinfo {author} {\bibfnamefont {G.~K.}\
		\bibnamefont {Batchelor}},\ }\href@noop {} {\emph {\bibinfo {title} {The
			Theory of Homogeneous Turbulence}}}\ (\bibinfo  {publisher} {Cambridge
	university press},\ \bibinfo {year} {1953})\BibitemShut {NoStop}%
\bibitem [{\citenamefont {Batchelor}\ and\ \citenamefont
	{Townsend}(1947)}]{Batchelor1947}%
\BibitemOpen
\bibfield  {author} {\bibinfo {author} {\bibfnamefont {G.~K.}~\bibnamefont
		{Batchelor}}\ and\ \bibinfo {author} {\bibfnamefont {A.~A.}~\bibnamefont
		{Townsend}},\ }\bibfield  {title} {\textit {\bibinfo {title} {Decay of
			Vorticity in Isotropic Turbulence},}\ }in\ \href@noop {} {\emph {\bibinfo
		{booktitle} {Proc. R. Soc. London Ser. A}}},\ Vol.\ \bibinfo {volume} {190}\
(\bibinfo {organization} {The Royal Society},\ \bibinfo {year} {1947})\ pp.\
\bibinfo {pages} {534--550}\BibitemShut {NoStop}%
\bibitem [{\citenamefont {Batchelor}\ and\ \citenamefont
	{Townsend}(1948{\natexlab{a}})}]{Batchelor1948a}%
\BibitemOpen
\bibfield  {author} {\bibinfo {author} {\bibfnamefont {G.~K.}~\bibnamefont
		{Batchelor}}\ and\ \bibinfo {author} {\bibfnamefont {A.~A.}~\bibnamefont
		{Townsend}},\ }\bibfield  {title} {\textit {\bibinfo {title} {Decay of
			Isotropic Turbulence in the Initial Period},}\ }in\ \href@noop {} {\emph
	{\bibinfo {booktitle} {Proc. R. Soc. London Ser. A}}},\ Vol.\ \bibinfo
{volume} {193}\ (\bibinfo {organization} {The Royal Society},\ \bibinfo
{year} {1948})\ pp.\ \bibinfo {pages} {539--558}\BibitemShut {NoStop}%
\bibitem [{\citenamefont {Batchelor}\ and\ \citenamefont
	{Townsend}(1948{\natexlab{b}})}]{Batchelor1948b}%
\BibitemOpen
\bibfield  {author} {\bibinfo {author} {\bibfnamefont {G.~K.}~\bibnamefont
		{Batchelor}}\ and\ \bibinfo {author} {\bibfnamefont {A.~A.}~\bibnamefont
		{Townsend}},\ }\bibfield  {title} {\textit {\bibinfo {title} {Decay of
			Turbulence in the Final Period},}\ }in\ \href@noop {} {\emph {\bibinfo
		{booktitle} {Proc. Roy. Soc. London Ser. A}}},\ Vol.\ \bibinfo {volume}
{194}\ (\bibinfo {organization} {The Royal Society},\ \bibinfo {year}
{1948})\ pp.\ \bibinfo {pages} {527--543}\BibitemShut {NoStop}%
\bibitem [{\citenamefont {Politano}\ and\ \citenamefont
	{Pouquet}(1998)}]{Politano1998PRE}%
\BibitemOpen
\bibfield  {author} {\bibinfo {author} {\bibfnamefont {H.}~\bibnamefont
		{Politano}}\ and\ \bibinfo {author} {\bibfnamefont {A.}~\bibnamefont
		{Pouquet}},\ }\bibfield  {title} {\textit {\bibinfo {title} {von K\'arm\'an
			Howarth Equation for Magnetohydrodynamics and its Consequences on Third-Order
			Longitudinal Structure and Correlation Functions},}\ }\href {\doibase
	10.1103/PhysRevE.57.R21} {\bibfield  {journal} {\bibinfo  {journal} {Phys.
			Rev. E}\ }\textbf {\bibinfo {volume} {57}},\ \bibinfo {pages} {R21--R24}
	(\bibinfo {year} {1998})}\BibitemShut {NoStop}%
\bibitem [{\citenamefont {Wan}\ \emph {et~al.}(2012)\citenamefont {Wan},
	\citenamefont {Oughton}, \citenamefont {Servidio},\ and\ \citenamefont
	{Matthaeus}}]{Wan2012JFM}%
\BibitemOpen
\bibfield  {author} {\bibinfo {author} {\bibfnamefont {M.}~\bibnamefont
		{Wan}}, \bibinfo {author} {\bibfnamefont {S.}~\bibnamefont {Oughton}},
	\bibinfo {author} {\bibfnamefont {S.}~\bibnamefont {Servidio}}, \ and\
	\bibinfo {author} {\bibfnamefont {W.~H.}\ \bibnamefont {Matthaeus}},\
}\bibfield  {title} {\textit {\bibinfo {title} {von K\'am\'an
		Self-Preservation Hypothesis for Magnetohydrodynamic Turbulence and its
		Consequences for Universality},}\ }\href {\doibase 10.1017/jfm.2012.61}
{\bibfield  {journal} {\bibinfo  {journal} {J. Fluid Mech.}\ }\textbf
	{\bibinfo {volume} {697}},\ \bibinfo {pages} {296–315} (\bibinfo {year}
	{2012})}\BibitemShut {NoStop}%
\bibitem [{\citenamefont {Wu}\ \emph {et~al.}(2013)\citenamefont {Wu},
	\citenamefont {Wan}, \citenamefont {Matthaeus}, \citenamefont {Shay},\ and\
	\citenamefont {Swisdak}}]{Wu2013PRL}%
\BibitemOpen
\bibfield  {author} {\bibinfo {author} {\bibfnamefont {P.}~\bibnamefont
		{Wu}}, \bibinfo {author} {\bibfnamefont {M.}~\bibnamefont {Wan}}, \bibinfo
	{author} {\bibfnamefont {W.~H.}\ \bibnamefont {Matthaeus}}, \bibinfo {author}
	{\bibfnamefont {M.~A.}\ \bibnamefont {Shay}}, \ and\ \bibinfo {author}
	{\bibfnamefont {M.}~\bibnamefont {Swisdak}},\ }\bibfield  {title} {\textit
	{\bibinfo {title} {von K\'arm\'an Energy Decay and Heating of Protons and
			Electrons in a Kinetic Turbulent Plasma},}\ }\href {\doibase
	10.1103/PhysRevLett.111.121105} {\bibfield  {journal} {\bibinfo  {journal}
		{Phys. Rev. Lett.}\ }\textbf {\bibinfo {volume} {111}},\ \bibinfo {pages}
	{121105} (\bibinfo {year} {2013})}\BibitemShut {NoStop}%
\bibitem [{\citenamefont {Parashar}\ \emph {et~al.}(2015)\citenamefont
	{Parashar}, \citenamefont {Matthaeus}, \citenamefont {Shay},\ and\
	\citenamefont {Wan}}]{Parashar:ApJL2015}%
\BibitemOpen
\bibfield  {author} {\bibinfo {author} {\bibfnamefont {T.~N.}\
		\bibnamefont {Parashar}}, \bibinfo {author} {\bibfnamefont {W.~H.}\
		\bibnamefont {Matthaeus}}, \bibinfo {author} {\bibfnamefont {M.~A.}\
		\bibnamefont {Shay}}, \ and\ \bibinfo {author} {\bibfnamefont {M.}\
		\bibnamefont {Wan}},\ }\bibfield  {title} {\textit {\bibinfo {title}
		{Transition from Kinetic to MHD Behavior in a Collisionless Plasma},}\ }\href
{http://stacks.iop.org/0004-637X/811/i=2/a=112} {\bibfield  {journal}
	{\bibinfo  {journal} {Astrophys. J.}\ }\textbf {\bibinfo {volume} {811}},\
	\bibinfo {pages} {112} (\bibinfo {year} {2015})}\BibitemShut {NoStop}%
\bibitem [{\citenamefont {Pearson}\ \emph {et~al.}(2002)\citenamefont
	{Pearson}, \citenamefont {Krogstad},\ and\ \citenamefont {van~de
		Water}}]{Pearson2002PoF}%
\BibitemOpen
\bibfield  {author} {\bibinfo {author} {\bibfnamefont {B.~R.}\ \bibnamefont
		{Pearson}}, \bibinfo {author} {\bibfnamefont {P.~A.}\ \bibnamefont
		{Krogstad}}, \ and\ \bibinfo {author} {\bibfnamefont {W.}~\bibnamefont
		{van~de Water}},\ }\bibfield  {title} {\textit {\bibinfo {title}
		{Measurements of the Turbulent Energy Dissipation Rate},}\ }\href {\doibase
	10.1063/1.1445422} {\bibfield  {journal} {\bibinfo  {journal} {Phys. Fluids}\
	}\textbf {\bibinfo {volume} {14}},\ \bibinfo {pages} {1288--1290} (\bibinfo
	{year} {2002})}\BibitemShut {NoStop}%
\bibitem [{\citenamefont {Kaneda}\ \emph {et~al.}(2003)\citenamefont {Kaneda},
	\citenamefont {Ishihara}, \citenamefont {Yokokawa}, \citenamefont {Itakura},\
	and\ \citenamefont {Uno}}]{Kaneda2003PoF}%
\BibitemOpen
\bibfield  {author} {\bibinfo {author} {\bibfnamefont {Y.}\ \bibnamefont
		{Kaneda}}, \bibinfo {author} {\bibfnamefont {T.}\ \bibnamefont
		{Ishihara}}, \bibinfo {author} {\bibfnamefont {M.}\ \bibnamefont
		{Yokokawa}}, \bibinfo {author} {\bibfnamefont {K.}\ \bibnamefont
		{Itakura}}, \ and\ \bibinfo {author} {\bibfnamefont {A.}\ \bibnamefont
		{Uno}},\ }\bibfield  {title} {\textit {\bibinfo {title} {Energy Dissipation
			Rate and Energy Spectrum in High Resolution Direct Numerical Simulations of
			Turbulence in a Periodic Box},}\ }\href {\doibase 10.1063/1.1539855}
{\bibfield  {journal} {\bibinfo  {journal} {Phys. Fluids}\ }\textbf
	{\bibinfo {volume} {15}},\ \bibinfo {pages} {L21--L24} (\bibinfo {year}
	{2003})}\BibitemShut {NoStop}%
\bibitem [{\citenamefont {Pearson}\ \emph {et~al.}(2004)\citenamefont
	{Pearson}, \citenamefont {Yousef}, \citenamefont {Haugen}, \citenamefont
	{Brandenburg},\ and\ \citenamefont {Krogstad}}]{Pearson2004PoF}%
\BibitemOpen
\bibfield  {author} {\bibinfo {author} {\bibfnamefont {B.~R.}\
		\bibnamefont {Pearson}}, \bibinfo {author} {\bibfnamefont {T.~A.}\
		\bibnamefont {Yousef}}, \bibinfo {author} {\bibfnamefont {N.~L.}\
		\bibnamefont {Haugen}}, \bibinfo {author} {\bibfnamefont {A.}\ \bibnamefont
		{Brandenburg}}, \ and\ \bibinfo {author} {\bibfnamefont {P.}\ \bibnamefont
		{Krogstad}},\ }\bibfield  {title} {\textit {\bibinfo {title} {Delayed
			Correlation Between Turbulent Energy Injection and Dissipation},}\ }\href
{\doibase 10.1103/PhysRevE.70.056301} {\bibfield  {journal} {\bibinfo
		{journal} {Phys. Rev. E}\ }\textbf {\bibinfo {volume} {70}},\ \bibinfo
	{pages} {056301} (\bibinfo {year} {2004})}\BibitemShut {NoStop}%
\bibitem [{\citenamefont {Usmanov}\ \emph {et~al.}(2014)\citenamefont
	{Usmanov}, \citenamefont {Goldstein},\ and\ \citenamefont
	{Matthaeus}}]{Usmanov2014ApJ}%
\BibitemOpen
\bibfield  {author} {\bibinfo {author} {\bibfnamefont {A.~V.}\
		\bibnamefont {Usmanov}}, \bibinfo {author} {\bibfnamefont {M.~L.}\
		\bibnamefont {Goldstein}}, \ and\ \bibinfo {author} {\bibfnamefont
		{W.~H.}\ \bibnamefont {Matthaeus}},\ }\bibfield  {title} {\textit
	{\bibinfo {title} {Three-Fluid, Three-Dimensional Magnetohydrodynamic Solar
			Wind Model with Eddy Viscosity and Turbulent Resistivity},}\ }\href
{http://stacks.iop.org/0004-637X/788/i=1/a=43} {\bibfield  {journal}
	{\bibinfo  {journal} {Astrophys. J.}\ }\textbf {\bibinfo {volume} {788}},\
	\bibinfo {pages} {43} (\bibinfo {year} {2014})}\BibitemShut {NoStop}%
\bibitem [{\citenamefont {Dallas}\ and\ \citenamefont
	{Alexakis}(2014)}]{Dallas2014ApJ}%
\BibitemOpen
\bibfield  {author} {\bibinfo {author} {\bibfnamefont {V.}~\bibnamefont
		{Dallas}}\ and\ \bibinfo {author} {\bibfnamefont {A.}~\bibnamefont
		{Alexakis}},\ }\bibfield  {title} {\textit {\bibinfo {title} {The Signature
			of Initial Conditions on Magnetohydrodynamic Turbulence},}\ }\href
{http://stacks.iop.org/2041-8205/788/i=2/a=L36} {\bibfield  {journal}
	{\bibinfo  {journal} {Astrophys. J. Lett.}\ }\textbf {\bibinfo {volume}
		{788}},\ \bibinfo {pages} {L36} (\bibinfo {year} {2014})}\BibitemShut
{NoStop}%
\bibitem [{\citenamefont {Linkmann}\ \emph {et~al.}(2015)\citenamefont
	{Linkmann}, \citenamefont {Berera}, \citenamefont {McComb},\ and\
	\citenamefont {McKay}}]{Linkmann2015PRL}%
\BibitemOpen
\bibfield  {author} {\bibinfo {author} {\bibfnamefont {M.~F.}\ \bibnamefont
		{Linkmann}}, \bibinfo {author} {\bibfnamefont {A.}~\bibnamefont {Berera}},
	\bibinfo {author} {\bibfnamefont {W.~D.}\ \bibnamefont {McComb}}, \ and\
	\bibinfo {author} {\bibfnamefont {M.~E.}\ \bibnamefont {McKay}},\ }\bibfield
{title} {\textit {\bibinfo {title} {Nonuniversality and Finite Dissipation
			in Decaying Magnetohydrodynamic Turbulence},}\ }\href {\doibase
	10.1103/PhysRevLett.114.235001} {\bibfield  {journal} {\bibinfo  {journal}
		{Phys. Rev. Lett.}\ }\textbf {\bibinfo {volume} {114}},\ \bibinfo {pages}
	{235001} (\bibinfo {year} {2015})}\BibitemShut {NoStop}%
\bibitem [{\citenamefont {Linkmann}\ \emph {et~al.}(2017)\citenamefont
	{Linkmann}, \citenamefont {Berera},\ and\ \citenamefont
	{Goldstraw}}]{Linkmann2017PRE}%
\BibitemOpen
\bibfield  {author} {\bibinfo {author} {\bibfnamefont {M.}\ \bibnamefont
		{Linkmann}}, \bibinfo {author} {\bibfnamefont {A.}\ \bibnamefont
		{Berera}}, \ and\ \bibinfo {author} {\bibfnamefont {E.~E.}\ \bibnamefont
		{Goldstraw}},\ }\bibfield  {title} {\textit {\bibinfo {title}
		{Reynolds-number Dependence of the Dimensionless Dissipation Rate in
			Homogeneous Magnetohydrodynamic Turbulence},}\ }\href {\doibase
	10.1103/PhysRevE.95.013102} {\bibfield  {journal} {\bibinfo  {journal} {Phys.
			Rev. E}\ }\textbf {\bibinfo {volume} {95}},\ \bibinfo {pages} {013102}
	(\bibinfo {year} {2017})}\BibitemShut {NoStop}%
\bibitem [{\citenamefont {McComb}\ and\ \citenamefont
	{Fairhurst}(2017)}]{McComb2017Arxiv}%
\BibitemOpen
\bibfield  {author} {\bibinfo {author} {\bibfnamefont {W.~D.}~\bibnamefont
		{McComb}}\ and\ \bibinfo {author} {\bibfnamefont {R.~B.}~\bibnamefont
		{Fairhurst}},\ }\bibfield  {title} {\textit {\bibinfo {title} {The
			Dimensionless Dissipation Rate and the Kolmogorov (1941) Hypothesis of Local
			Stationarity in Freely Decaying Isotropic Turbulence},}\ }\href@noop {}
{\bibfield  {journal} {\bibinfo  {journal} {arXiv preprint arXiv:1712.05300}\
	} (\bibinfo {year} {2017})}\BibitemShut {NoStop}%
\bibitem [{\citenamefont {Hossain}\ \emph {et~al.}(1995)\citenamefont
	{Hossain}, \citenamefont {Gray}, \citenamefont {Pontius}, \citenamefont
	{Matthaeus},\ and\ \citenamefont {Oughton}}]{Hossain1995PoP}%
\BibitemOpen
\bibfield  {author} {\bibinfo {author} {\bibfnamefont {M.}~\bibnamefont
		{Hossain}}, \bibinfo {author} {\bibfnamefont {P.~C.}\ \bibnamefont {Gray}},
	\bibinfo {author} {\bibfnamefont {D.~H.}\ \bibnamefont {Pontius}}, \bibinfo
	{author} {\bibfnamefont {W.~H.}\ \bibnamefont {Matthaeus}}, \ and\ \bibinfo
	{author} {\bibfnamefont {S.}~\bibnamefont {Oughton}},\ }\bibfield  {title}
{\textit {\bibinfo {title} {Phenomenology for the Decay of
			Energy-Containing Eddies in Homogeneous MHD Turbulence},}\ }\href {\doibase
	10.1063/1.868665} {\bibfield  {journal} {\bibinfo  {journal} {Phys. Fluids}\
	}\textbf {\bibinfo {volume} {7}},\ \bibinfo {pages} {2886--2904} (\bibinfo
	{year} {1995})}\BibitemShut {NoStop}%
\bibitem [{\citenamefont {Hossain}\ \emph {et~al.}(1996)\citenamefont
	{Hossain}, \citenamefont {Gray}, \citenamefont {Jr.}, \citenamefont
	{Matthaeus},\ and\ \citenamefont {Oughton}}]{Hossain1996AIP}%
\BibitemOpen
\bibfield  {author} {\bibinfo {author} {\bibfnamefont {M.}\ \bibnamefont
		{Hossain}}, \bibinfo {author} {\bibfnamefont {P.~C.}\ \bibnamefont
		{Gray}}, \bibinfo {author} {\bibfnamefont {D.~H.~Pontius}\ \bibnamefont
		{Jr.}}, \bibinfo {author} {\bibfnamefont {W.~H.}\ \bibnamefont
		{Matthaeus}}, \ and\ \bibinfo {author} {\bibfnamefont {S.}\ \bibnamefont
		{Oughton}},\ }\bibfield  {title} {\textit {\bibinfo {title} {Is the
			Alfv\'en-Wave Propagation Effect Important for Energy Decay in Homogeneous MHD
			Turbulence?}}\ }\href {\doibase 10.1063/1.51471} {\bibfield  {journal}
	{\bibinfo  {journal} {AIP Conf. Proc.}\ }\textbf {\bibinfo
		{volume} {382}},\ \bibinfo {pages} {358--361} (\bibinfo {year}
	{1996})}\BibitemShut {NoStop}%
\bibitem [{\citenamefont {Bigot}\ \emph
	{et~al.}(2008{\natexlab{a}})\citenamefont {Bigot}, \citenamefont {Galtier},\
	and\ \citenamefont {Politano}}]{Bigot2008PRL}%
\BibitemOpen
\bibfield  {author} {\bibinfo {author} {\bibfnamefont {B.}\ \bibnamefont
		{Bigot}}, \bibinfo {author} {\bibfnamefont {S.}\ \bibnamefont
		{Galtier}}, \ and\ \bibinfo {author} {\bibfnamefont {H.}\
		\bibnamefont {Politano}},\ }\bibfield  {title} {\textit {\bibinfo {title}
		{Energy Decay Laws in Strongly Anisotropic Magnetohydrodynamic Turbulence},}\
}\href {\doibase 10.1103/PhysRevLett.100.074502} {\bibfield  {journal}
{\bibinfo  {journal} {Phys. Rev. Lett.}\ }\textbf {\bibinfo {volume} {100}},\
\bibinfo {pages} {074502} (\bibinfo {year} {2008}{\natexlab{a}})}\BibitemShut
{NoStop}%
\bibitem [{\citenamefont {Bigot}\ \emph
	{et~al.}(2008{\natexlab{b}})\citenamefont {Bigot}, \citenamefont {Galtier},\
	and\ \citenamefont {Politano}}]{Bigot2008PRE}%
\BibitemOpen
\bibfield  {author} {\bibinfo {author} {\bibfnamefont {B.}\ \bibnamefont
		{Bigot}}, \bibinfo {author} {\bibfnamefont {S.}\ \bibnamefont
		{Galtier}}, \ and\ \bibinfo {author} {\bibfnamefont {H.}\
		\bibnamefont {Politano}},\ }\bibfield  {title} {\textit {\bibinfo {title}
		{Development of Anisotropy in Incompressible Magnetohydrodynamic
			Turbulence},}\ }\href {\doibase 10.1103/PhysRevE.78.066301} {\bibfield
	{journal} {\bibinfo  {journal} {Phys. Rev. E}\ }\textbf {\bibinfo {volume}
		{78}},\ \bibinfo {pages} {066301} (\bibinfo {year}
	{2008}{\natexlab{b}})}\BibitemShut {NoStop}%
\bibitem [{\citenamefont {Bigot}\ and\ \citenamefont
	{Galtier}(2011)}]{Bigot2011PRE}%
\BibitemOpen
\bibfield  {author} {\bibinfo {author} {\bibfnamefont {B.}\ \bibnamefont
		{Bigot}}\ and\ \bibinfo {author} {\bibfnamefont {S.}\ \bibnamefont
		{Galtier}},\ }\bibfield  {title} {\textit {\bibinfo {title} {Two-Dimensional
			State in Driven Magnetohydrodynamic Turbulence},}\ }\href {\doibase
	10.1103/PhysRevE.83.026405} {\bibfield  {journal} {\bibinfo  {journal} {Phys.
			Rev. E}\ }\textbf {\bibinfo {volume} {83}},\ \bibinfo {pages} {026405}
	(\bibinfo {year} {2011})}\BibitemShut {NoStop}%
\bibitem [{\citenamefont {Zhdankin}\ \emph {et~al.}(2017)\citenamefont
	{Zhdankin}, \citenamefont {Boldyrev},\ and\ \citenamefont
	{Mason}}]{Zhdankin2017MNRAS}%
\BibitemOpen
\bibfield  {author} {\bibinfo {author} {\bibfnamefont {V.}\
		\bibnamefont {Zhdankin}}, \bibinfo {author} {\bibfnamefont {S.}\
		\bibnamefont {Boldyrev}}, \ and\ \bibinfo {author} {\bibfnamefont {J.}\
		\bibnamefont {Mason}},\ }\bibfield  {title} {\textit {\bibinfo {title}
		{Influence of a Large-Scale Field on Energy Dissipation in
			Magnetohydrodynamic Turbulence},}\ }\href {\doibase 10.1093/mnras/stx611}
{\bibfield  {journal} {\bibinfo  {journal} {Mon. Not. R.
			Astron. Soc.}\ }\textbf {\bibinfo {volume} {468}},\ \bibinfo {pages}
	{4025--4029} (\bibinfo {year} {2017})}\BibitemShut {NoStop}%
\bibitem [{\citenamefont {Shebalin}\ \emph {et~al.}(1983)\citenamefont
	{Shebalin}, \citenamefont {Matthaeus},\ and\ \citenamefont
	{Montgomery}}]{Shebalin1983JPP}%
\BibitemOpen
\bibfield  {author} {\bibinfo {author} {\bibfnamefont {J.~V.}\ \bibnamefont
		{Shebalin}}, \bibinfo {author} {\bibfnamefont {W.~H.}\ \bibnamefont
		{Matthaeus}}, \ and\ \bibinfo {author} {\bibfnamefont {D.}\ \bibnamefont
		{Montgomery}},\ }\bibfield  {title} {\textit {\bibinfo {title} {Anisotropy
			in MHD Turbulence Due to a Mean Magnetic Field},}\ }\href {\doibase
	10.1017/S0022377800000933} {\bibfield  {journal} {\bibinfo  {journal} {J.
			Plasma Phys.}\ }\textbf {\bibinfo {volume} {29}},\ \bibinfo {pages}
	{525–547} (\bibinfo {year} {1983})}\BibitemShut {NoStop}%
\bibitem [{\citenamefont {Oughton}\ \emph {et~al.}(1994)\citenamefont
	{Oughton}, \citenamefont {Priest},\ and\ \citenamefont
	{Matthaeus}}]{Oughton1994JFM}%
\BibitemOpen
\bibfield  {author} {\bibinfo {author} {\bibfnamefont {S.}\ \bibnamefont
		{Oughton}}, \bibinfo {author} {\bibfnamefont {E.~R.}\ \bibnamefont
		{Priest}}, \ and\ \bibinfo {author} {\bibfnamefont {W.~H.}\ \bibnamefont
		{Matthaeus}},\ }\bibfield  {title} {\textit {\bibinfo {title} {The Influence
			of a Mean Magnetic Field on Three-Dimensional Magnetohydrodynamic
			Turbulence},}\ }\href {\doibase 10.1017/S0022112094002867} {\bibfield
	{journal} {\bibinfo  {journal} {J. Fluid Mech.}\ }\textbf {\bibinfo {volume}
		{280}},\ \bibinfo {pages} {95–117} (\bibinfo {year} {1994})}\BibitemShut
{NoStop}%
\bibitem [{\citenamefont {Wan}\ \emph {et~al.}(2010)\citenamefont {Wan},
	\citenamefont {Oughton}, \citenamefont {Servidio},\ and\ \citenamefont
	{Matthaeus}}]{Wan2010bPoP}%
\BibitemOpen
\bibfield  {author} {\bibinfo {author} {\bibfnamefont {M.}~\bibnamefont
		{Wan}}, \bibinfo {author} {\bibfnamefont {S.}~\bibnamefont {Oughton}},
	\bibinfo {author} {\bibfnamefont {S.}~\bibnamefont {Servidio}}, \ and\
	\bibinfo {author} {\bibfnamefont {W.~H.}\ \bibnamefont {Matthaeus}},\
}\bibfield  {title} {\textit {\bibinfo {title} {On the Accuracy of
		Simulations of Turbulence},}\ }\href {\doibase 10.1063/1.3474957} {\bibfield
{journal} {\bibinfo  {journal} {Phys. Plasmas}\ }\textbf {\bibinfo {volume}
	{17}},\ \bibinfo {pages} {082308} (\bibinfo {year} {2010})}\BibitemShut
{NoStop}%
\bibitem [{\citenamefont {Zhou}\ \emph {et~al.}(2004)\citenamefont {Zhou},
	\citenamefont {Matthaeus},\ and\ \citenamefont {Dmitruk}}]{Zhou2004RMP}%
\BibitemOpen
\bibfield  {author} {\bibinfo {author} {\bibfnamefont {Y.}~\bibnamefont
		{Zhou}}, \bibinfo {author} {\bibfnamefont {W.~H.}\ \bibnamefont {Matthaeus}},
	\ and\ \bibinfo {author} {\bibfnamefont {P.}~\bibnamefont {Dmitruk}},\
}\bibfield  {title} {\textit {\bibinfo {title} {Colloquium:
		Magnetohydrodynamic Turbulence and Time Scales in Astrophysical and Space
		Plasmas},}\ }\href {\doibase 10.1103/RevModPhys.76.1015} {\bibfield
{journal} {\bibinfo  {journal} {Rev. Mod. Phys.}\ }\textbf {\bibinfo {volume}
	{76}},\ \bibinfo {pages} {1015--1035} (\bibinfo {year} {2004})}\BibitemShut
{NoStop}%
\bibitem [{\citenamefont {Oughton}\ \emph {et~al.}(2017)\citenamefont
	{Oughton}, \citenamefont {Matthaeus},\ and\ \citenamefont
	{Dmitruk}}]{Oughton2017ApJ}%
\BibitemOpen
\bibfield  {author} {\bibinfo {author} {\bibfnamefont {S.}~\bibnamefont
		{Oughton}}, \bibinfo {author} {\bibfnamefont {W.~H.}\ \bibnamefont
		{Matthaeus}}, \ and\ \bibinfo {author} {\bibfnamefont {P.}~\bibnamefont
		{Dmitruk}},\ }\bibfield  {title} {\textit {\bibinfo {title} {Reduced MHD in
			Astrophysical Applications: Two-Dimensional or Three-Dimensional?}}\ }\href
{http://stacks.iop.org/0004-637X/839/i=1/a=2} {\bibfield  {journal} {\bibinfo
		{journal} {Astrophys. J.}\ }\textbf {\bibinfo {volume} {839}},\ \bibinfo
	{pages} {2} (\bibinfo {year} {2017})}\BibitemShut {NoStop}%
\bibitem [{\citenamefont {Galtier}\ \emph {et~al.}(2000)\citenamefont
	{Galtier}, \citenamefont {Nazarenko}, \citenamefont {Newell},\ and\
	\citenamefont {Pouquet}}]{Galtier2000JPP}%
\BibitemOpen
\bibfield  {author} {\bibinfo {author} {\bibfnamefont {S.}~\bibnamefont
		{Galtier}}, \bibinfo {author} {\bibfnamefont {S.~V.}\ \bibnamefont
		{Nazarenko}}, \bibinfo {author} {\bibfnamefont {A.~C.}\ \bibnamefont
		{Newell}}, \ and\ \bibinfo {author} {\bibfnamefont {A.}~\bibnamefont
		{Pouquet}},\ }\bibfield  {title} {\textit {\bibinfo {title} {A Weak
			Turbulence Theory for Incompressible Magnetohydrodynamics},}\ }\href@noop {}
{\bibfield  {journal} {\bibinfo  {journal} {J. Plasma Phys.}\
	}\textbf {\bibinfo {volume} {63}},\ \bibinfo {pages} {447–488} (\bibinfo
	{year} {2000})}\BibitemShut {NoStop}%
\bibitem [{\citenamefont {Meyrand}\ \emph {et~al.}(2016)\citenamefont
	{Meyrand}, \citenamefont {Galtier},\ and\ \citenamefont
	{Kiyani}}]{Meyrand2016PRL}%
\BibitemOpen
\bibfield  {author} {\bibinfo {author} {\bibfnamefont {R.}\ \bibnamefont
		{Meyrand}}, \bibinfo {author} {\bibfnamefont {S.}\ \bibnamefont
		{Galtier}}, \ and\ \bibinfo {author} {\bibfnamefont {K.~H.}\ \bibnamefont
		{Kiyani}},\ }\bibfield  {title} {\textit {\bibinfo {title} {Direct Evidence
			of the Transition from Weak to Strong Magnetohydrodynamic Turbulence},}\
}\href {\doibase 10.1103/PhysRevLett.116.105002} {\bibfield  {journal}
{\bibinfo  {journal} {Phys. Rev. Lett.}\ }\textbf {\bibinfo {volume} {116}},\
\bibinfo {pages} {105002} (\bibinfo {year} {2016})}\BibitemShut {NoStop}%
\bibitem [{\citenamefont {Dmitruk}\ and\ \citenamefont
	{Matthaeus}(2009)}]{Dmitruk:PoP2009}%
\BibitemOpen
\bibfield  {author} {\bibinfo {author} {\bibfnamefont {P.}~\bibnamefont
		{Dmitruk}}\ and\ \bibinfo {author} {\bibfnamefont {W.~H.}\ \bibnamefont
		{Matthaeus}},\ }\bibfield  {title} {\textit {\bibinfo {title} {Waves and
			Turbulence in Magnetohydrodynamic Direct Numerical Simulations},}\ }\href
{\doibase 10.1063/1.3148335} {\bibfield  {journal} {\bibinfo  {journal}
		{Phys. Plasmas}\ }\textbf {\bibinfo {volume} {16}},\ \bibinfo {pages}
	{062304} (\bibinfo {year} {2009})}\BibitemShut {NoStop}%
\bibitem [{\citenamefont {Breech}\ \emph {et~al.}(2008)\citenamefont {Breech},
	\citenamefont {Matthaeus}, \citenamefont {Minnie}, \citenamefont {Bieber},
	\citenamefont {Oughton}, \citenamefont {Smith},\ and\ \citenamefont
	{Isenberg}}]{Breech2008JGR}%
\BibitemOpen
\bibfield  {author} {\bibinfo {author} {\bibfnamefont {B.}~\bibnamefont
		{Breech}}, \bibinfo {author} {\bibfnamefont {W.~H.}\ \bibnamefont
		{Matthaeus}}, \bibinfo {author} {\bibfnamefont {J.}~\bibnamefont {Minnie}},
	\bibinfo {author} {\bibfnamefont {J.~W.}\ \bibnamefont {Bieber}}, \bibinfo
	{author} {\bibfnamefont {S.}~\bibnamefont {Oughton}}, \bibinfo {author}
	{\bibfnamefont {C.~W.}\ \bibnamefont {Smith}}, \ and\ \bibinfo {author}
	{\bibfnamefont {P.~A.}\ \bibnamefont {Isenberg}},\ }\bibfield  {title}
{\textit {\bibinfo {title} {Turbulence Transport Throughout the
			Heliosphere},}\ }\href {\doibase 10.1029/2007JA012711} {\bibfield  {journal}
	{\bibinfo  {journal} {J. Geophys. Res.}\ }\textbf
	{\bibinfo {volume} {113}} (\bibinfo {year} {2008}),\
	10.1029/2007JA012711}\BibitemShut {NoStop}%
\bibitem [{\citenamefont {MacBride}\ \emph {et~al.}(2008)\citenamefont
	{MacBride}, \citenamefont {Smith},\ and\ \citenamefont
	{Forman}}]{MacBride2008ApJ}%
\BibitemOpen
\bibfield  {author} {\bibinfo {author} {\bibfnamefont {Benjamin~T.}\
		\bibnamefont {MacBride}}, \bibinfo {author} {\bibfnamefont {Charles~W.}\
		\bibnamefont {Smith}}, \ and\ \bibinfo {author} {\bibfnamefont {Miriam~A.}\
		\bibnamefont {Forman}},\ }\bibfield  {title} {\textit {\bibinfo {title} {The
			Turbulent Cascade at 1 AU: Energy Transfer and the Third-Order Scaling for
			MHD},}\ }\href {\doibase 10.1086/529575} {\bibfield  {journal} {\bibinfo
		{journal} {Astrophys. J.}\ }\textbf {\bibinfo {volume} {679}},\ \bibinfo
	{pages} {1644} (\bibinfo {year} {2008})}\BibitemShut {NoStop}%
\bibitem [{\citenamefont {Sorriso-Valvo}\ \emph {et~al.}(2007)\citenamefont
	{Sorriso-Valvo}, \citenamefont {Marino}, \citenamefont {Carbone},
	\citenamefont {Noullez}, \citenamefont {Lepreti}, \citenamefont {Veltri},
	\citenamefont {Bruno}, \citenamefont {Bavassano},\ and\ \citenamefont
	{Pietropaolo}}]{Sorriso-Valvo2007PRL}%
\BibitemOpen
\bibfield  {author} {\bibinfo {author} {\bibfnamefont {L.}~\bibnamefont
		{Sorriso-Valvo}}, \bibinfo {author} {\bibfnamefont {R.}~\bibnamefont
		{Marino}}, \bibinfo {author} {\bibfnamefont {V.}~\bibnamefont {Carbone}},
	\bibinfo {author} {\bibfnamefont {A.}~\bibnamefont {Noullez}}, \bibinfo
	{author} {\bibfnamefont {F.}~\bibnamefont {Lepreti}}, \bibinfo {author}
	{\bibfnamefont {P.}~\bibnamefont {Veltri}}, \bibinfo {author} {\bibfnamefont
		{R.}~\bibnamefont {Bruno}}, \bibinfo {author} {\bibfnamefont
		{B.}~\bibnamefont {Bavassano}}, \ and\ \bibinfo {author} {\bibfnamefont
		{E.}~\bibnamefont {Pietropaolo}},\ }\bibfield  {title} {\textit {\bibinfo
		{title} {Observation of Inertial Energy Cascade in Interplanetary Space
			Plasma},}\ }\href {\doibase 10.1103/PhysRevLett.99.115001} {\bibfield
	{journal} {\bibinfo  {journal} {Phys. Rev. Lett.}\ }\textbf {\bibinfo
		{volume} {99}},\ \bibinfo {pages} {115001} (\bibinfo {year}
	{2007})}\BibitemShut {NoStop}%
\bibitem [{\citenamefont {Coburn}\ \emph {et~al.}(2015)\citenamefont {Coburn},
	\citenamefont {Forman}, \citenamefont {Smith}, \citenamefont {Vasquez},\ and\
	\citenamefont {Stawarz}}]{Coburn2014PRS}%
\BibitemOpen
\bibfield  {author} {\bibinfo {author} {\bibfnamefont {J.~T.}\
		\bibnamefont {Coburn}}, \bibinfo {author} {\bibfnamefont {M.~A.}\
		\bibnamefont {Forman}}, \bibinfo {author} {\bibfnamefont {C.~W.}\
		\bibnamefont {Smith}}, \bibinfo {author} {\bibfnamefont {B.~J.}\
		\bibnamefont {Vasquez}}, \ and\ \bibinfo {author} {\bibfnamefont {J.~E.}\
		\bibnamefont {Stawarz}},\ }\bibfield  {title} {\textit {\bibinfo {title}
		{Third-Moment Descriptions of the Interplanetary Turbulent Cascade,
			Intermittency and Back Transfer},}\ }\href {\doibase 10.1098/rsta.2014.0150}
{\bibfield  {journal} {\bibinfo  {journal} {Phil. Trans. R.
			Soc. A}\ }\textbf
	{\bibinfo {volume} {373}} (\bibinfo {year} {2015})}\BibitemShut {NoStop}%
\bibitem [{\citenamefont {van~der Holst}\ \emph {et~al.}(2014)\citenamefont
	{van~der Holst}, \citenamefont {Sokolov}, \citenamefont {Meng}, \citenamefont
	{Jin}, \citenamefont {Manchester}, \citenamefont {Toth},\ and\ \citenamefont
	{Gombosi}}]{vanderHolst2014ApJ}%
\BibitemOpen
\bibfield  {author} {\bibinfo {author} {\bibfnamefont {B.}~\bibnamefont
		{van~der Holst}}, \bibinfo {author} {\bibfnamefont {I.~V.}\ \bibnamefont
		{Sokolov}}, \bibinfo {author} {\bibfnamefont {X.}~\bibnamefont {Meng}},
	\bibinfo {author} {\bibfnamefont {M.}~\bibnamefont {Jin}}, \bibinfo {author}
	{\bibfnamefont {W.~B.}\ \bibnamefont {Manchester}, \bibfnamefont {IV}},
	\bibinfo {author} {\bibfnamefont {G.}~\bibnamefont {Toth}}, \ and\ \bibinfo
	{author} {\bibfnamefont {T.~I.}\ \bibnamefont {Gombosi}},\ }\bibfield
{title} {\textit {\bibinfo {title} {Alfv\'en Wave Solar Model (AWSoM): Coronal
			Heating},}\ }\href {\doibase 10.1088/0004-637X/782/2/81} {\bibfield
	{journal} {\bibinfo  {journal} {Astrophys. J.}\ }\textbf
	{\bibinfo {volume} {782}},\ \bibinfo {pages} {81} (\bibinfo {year}
	{2014})}\BibitemShut {NoStop}%
\bibitem [{\citenamefont {Lionello}\ \emph {et~al.}(2014)\citenamefont
	{Lionello}, \citenamefont {Velli}, \citenamefont {Downs}, \citenamefont
	{Linker}, \citenamefont {Mikic},\ and\ \citenamefont
	{Verdini}}]{Lionello2014ApJ}%
\BibitemOpen
\bibfield  {author} {\bibinfo {author} {\bibfnamefont {R.}~\bibnamefont
		{Lionello}}, \bibinfo {author} {\bibfnamefont {M.}~\bibnamefont {Velli}},
	\bibinfo {author} {\bibfnamefont {C.}~\bibnamefont {Downs}}, \bibinfo
	{author} {\bibfnamefont {J.~A.}\ \bibnamefont {Linker}}, \bibinfo {author}
	{\bibfnamefont {Z.}~\bibnamefont {Mikic}}, \ and\ \bibinfo {author}
	{\bibfnamefont {A.}~\bibnamefont {Verdini}},\ }\bibfield  {title} {\textit
	{\bibinfo {title} {Validating a Time-Dependent Turbulence-Driven Model of the
			Solar Wind},}\ }\href {\doibase 10.1088/0004-637X/784/2/120} {\bibfield
	{journal} {\bibinfo  {journal} {Astrophys. J.}\ }\textbf
	{\bibinfo {volume} {784}},\ \bibinfo {pages} {120} (\bibinfo {year}
	{2014})}\BibitemShut {NoStop}%
\bibitem [{\citenamefont {Smith}\ \emph {et~al.}(2004)\citenamefont {Smith},
	\citenamefont {Ghosh}, \citenamefont {Dmitruk},\ and\ \citenamefont
	{Matthaeus}}]{Smith2004GRL}%
\BibitemOpen
\bibfield  {author} {\bibinfo {author} {\bibfnamefont {D.}~\bibnamefont
		{Smith}}, \bibinfo {author} {\bibfnamefont {S.}~\bibnamefont {Ghosh}},
	\bibinfo {author} {\bibfnamefont {P.}~\bibnamefont {Dmitruk}}, \ and\
	\bibinfo {author} {\bibfnamefont {W.~H.}\ \bibnamefont {Matthaeus}},\
}\bibfield  {title} {\textit {\bibinfo {title} {Hall and Turbulence Effects
		on Magnetic Reconnection},}\ }\href {\doibase 10.1029/2003GL018689}
{\bibfield  {journal} {\bibinfo  {journal} {Geophys. R. Lett.}\
	}\textbf {\bibinfo {volume} {31}} (\bibinfo {year} {2004})}\BibitemShut {NoStop}%
\bibitem [{\citenamefont {Zhou}(1995)}]{Zhou1995PoF}%
\BibitemOpen
\bibfield  {author} {\bibinfo {author} {\bibfnamefont {Y.}~\bibnamefont
		{Zhou}},\ }\bibfield  {title} {\textit {\bibinfo {title} {A Phenomenological
			Treatment of Rotating Turbulence},}\ }\href {\doibase 10.1063/1.868457}
{\bibfield  {journal} {\bibinfo  {journal} {Phys. Fluids}\ }\textbf
	{\bibinfo {volume} {7}},\ \bibinfo {pages} {2092--2094} (\bibinfo {year}
	{1995})}\BibitemShut {NoStop}%
\end{thebibliography}

%

\end{document}